\documentclass[12pt,a4]{article}

\usepackage{amsfonts}
\usepackage{amstext}
\usepackage{amsmath}
\usepackage{amssymb}

\usepackage{mathrsfs}

\usepackage{graphicx}
\usepackage{color}
\usepackage{here}

\usepackage{ascmac}

\usepackage{mathtools}

\usepackage{ulem}
\DeclareRobustCommand{\erase}{\bgroup\markoverwith{\textcolor{red}{\rule[.5ex]{2pt}{1.0pt}}}\ULon}

\usepackage{bm}
\usepackage[dvipsnames]{xcolor}

\newcommand{\rf}[1]{(\ref{#1})}
\newcommand{\beq}{\begin{equation}}
\newcommand{\eeq}{\end{equation}}
\newcommand{\bea}{\begin{eqnarray}}
\newcommand{\eea}{\end{eqnarray}}

\allowdisplaybreaks[3]      

\newcommand{\e}{{\rm e}}

\newcommand{\g}{\gamma}

\renewcommand{\l}{\lambda}
\renewcommand{\L}{\Lambda}
\renewcommand{\b}{\beta}
\renewcommand{\a}{\alpha}

\newcommand{\dd}{{\rm d}}

\newcommand{\om}{\omega}
\newcommand{\del}{\delta}

\renewcommand{\k}{\kappa}

\newcommand{\oh}{\frac{1}{2}}

\newcommand{\dg}{\dagger}

\newcommand{\tr}{{\rm tr}}

\newcommand{\ra}{\rangle}
\newcommand{\la}{\langle}
\newcommand{\prt}{\partial}

\newcommand{\halftinyspace}{\hspace{0.0278em}} 
\newcommand{\tinyspace}{\hspace{0.0556em}}
\newcommand{\trehalftinyspace}{\hspace{0.0834em}}
\newcommand{\dbltinyspace}{\hspace{0.1112em}}

\newcommand{\negtinyspace}{\hspace{-0.0556em}}
\newcommand{\negdbltinyspace}{\hspace{-0.1112em}}
\newcommand{\negtrpltinyspace}{\hspace{-0.1668em}}
\newcommand{\negqntpltinyspace}{\hspace{-0.2780em}}

\newcommand{\define}{\leftdefine}
\newcommand{\leftdefine}{:=}

\font\twelvemsbm = msbm10 scaled\magstep1
\font\tenmsbm = msbm10
\font\eightmsbm = msbm8
\font\sixmsbm = msbm6
\newcommand{\Dbl}[1]{\leavevmode\raise-.10ex\hbox{\twelvemsbm #1}}
\newcommand{\dbl}[1]{\leavevmode\raise-.00ex\hbox{\tenmsbm #1}}
\newcommand{\dblsmall}[1]{\leavevmode\raise-.05ex\hbox{\eightmsbm #1}}
\newcommand{\dbltiny}[1]{\leavevmode\raise-.05ex\hbox{\sixmsbm #1}}

\newcommand{\half}{\frac{1}{2}}
\newcommand{\onethird}{\frac{1}{3}}

\def\void{}
\def\labelmark{}

\newenvironment{formula}[1]{\def\labelname{#1}
\ifx\void\labelname\def\junk{\begin{displaymath}}
\else\def\junk{\begin{equation}\label{\labelname}}\fi\junk}%
{\ifx\void\labelname\def\junk{\end{displaymath}}
\else\def\junk{\end{equation}}\fi\junk\labelmark\def\labelname{}}

{\ifx\void\labelname\def\junk{\end{array}\end{displaymath}}
\else\def\junk{\end{array}\right.\end{equation}}
\fi\junk\labelmark\def\labelname{}\def\junk{}
}

\newcommand{\beqv}{\begin{formula}{}}

\newcommand{\bra}[1]{\langle #1 |}

\newcommand{\ket}[1]{| #1 \rangle}

\newcommand{\vac}{\bra{{\rm vac}}}
\newcommand{\cuum}{\ket{{\rm vac}}}
\newcommand{\expect}[1]{\langle #1 \tinyspace\rangle}

\newcommand{\combi}[2]{\left( \!\! \begin{array}{c} 
	\raise0.5ex\hbox{$#1$} \\ \lower0.5ex\hbox{$#2$} \\ 
	\end{array} \!\! \right)}

\allowdisplaybreaks[4]	

\newcommand{\ScF}{a}

\newcommand{\DWplus}{\hbox{\hspace{-4pt}
  \textcolor{PineGreen}
            {\raise-0.27ex\hbox{\Large$\bm{+}$}}%
\hspace{-0pt}}}
\newcommand{\DWminus}{\hbox{\hspace{-4pt}
  \textcolor{PineGreen}
            {\raise-0.27ex\hbox{\Large$\bm{-}$}}%
\hspace{-0pt}}}
\newcommand{\DWpm}{\hbox{\hspace{-4pt}
  \textcolor{PineGreen}
            {\raise-0.27ex\hbox{\Large$\bm{\pm}$}}%
\hspace{-0pt}}}
\newcommand{\DWmp}{\hbox{\hspace{-4pt}
  \textcolor{PineGreen}
            {\raise-0.27ex\hbox{\Large$\bm{\mp}$}}%
\hspace{-0pt}}}

\newcommand{\DWgtrless}{\hbox{\hspace{-4pt}
  \textcolor{PineGreen}
            {\raise-0.27ex\hbox{\Large$\bm{\gtrless}$}}%
\hspace{-0pt}}}
\newcommand{\DWlessgtr}{\hbox{\hspace{-4pt}
  \textcolor{PineGreen}
            {\raise-0.27ex\hbox{\Large$\bm{\lessgtr}$}}%
\hspace{-0pt}}}

\newcommand{\hH}{{\hat{H}}}
\newcommand{\tH}{{\tilde{H}}}
\newcommand{\ha}{\hat{\alpha}}

\newcommand{\T}{t} 

\newcommand{\Length}{L}

\newcommand{\versionII}[1]{#1}

\begin{document}
\topmargin 0pt
\oddsidemargin 5mm
\headheight 0pt
\headsep 0pt
\topskip 9mm

\rightline{\today}

\begin{center}

\vspace{24pt}

{\Large \bf \versionII{The emergence of the  universe}}\\

\vspace{24pt}

{\sl J.\ Ambj\o rn}$\,^{a,b}$
and 
{\sl Y.\ Watabiki}$\,^{c}$

\vspace{10pt}

{\small

$^a$~The Niels Bohr Institute, Copenhagen University\\
Blegdamsvej 17, DK-2100 Copenhagen \O , Denmark.\\
email: ambjorn@nbi.dk

\vspace{10pt}

$^b$~Institute for Mathematics, Astrophysics and Particle Physics
(IMAPP)\\ \versionII{Radboud} University Nijmegen, Heyendaalseweg 135, 6525 AJ, \\
Nijmegen, The Netherlands

\vspace{10pt}

$^c$~Department of Physics, 
Institute of Science Tokyo,\\
Oh-okayama 2-12-1, Meguro-ku, Tokyo 152-8551, Japan\\
{email: yoshiyuki.watabiki@gmail.com}

}

\end{center}

\vspace{24pt}

\begin{center}
{\bf Abstract}
\end{center}

\vspace{6pt}

\noindent
We show how our Universe can emerge from a symmetry breaking of 
a multicomponent $W_3$ algebra, where  the components 
in addition form a Jordan algebra. We discuss how symmetry breaking 
related to the Jordan algebras $H_3(C)$ and $H_3(O)$ over the 
complex and octonion numbers can lead to an extended 
four-dimensional spacetime, where the expansion of the Universe is 
governed by a modified Friedmann equation. We finally discuss how this modified Friedmann equation might explain a number of puzzling cosmological observations.

\newpage

\section{Introduction}

We have presented models of our Universe based on an underlying theory 
 of Jordan algebras \cite{aw1,aw2,w1}. The automorphism groups of 
 the Jordan algebras suggested a picture not too dissimilar to the 
 picture one hopes for in string theory: that  one-dimensional universes 
 (in the case of string theory, strings) via interactions form the 
 present extended universe we observe today plus a compactified 
 space we have not yet observed. In the Jordan algebra case we 
 denoted this mechanism ``knitting'': due to the Jordan algebra structure
 certain of the one-dimensional universes were glued together, the interaction
 mediated via wormholes, themselves small one-dimensional universes.
 At distances larger that these small wormholes one would then observe 
 what could be viewed as higher dimensional spacetimes. We argued 
 that these higher dimensional spacetimes would satisfy a modified 
 Friedmann equation that describes the observed late time cosmology
 well \cite{mod1,mod2,mod3}. 
 
 \versionII{
While the underlying picture of ``emergence of a universe looking like ours'' 
has been discussed in detail in \cite{aw2}, 
let us nevertheless repeat aspects of this in the context of 
the ubiquitous ``emergence'' one encounters in theoretical physics. 
Underlying our theory is GCDT 
(generalized causal dynamical triangulations, 
see next section for a discussion). 
This is a modification of non-critical string theory, 
and like non-critical string theory it describes 
how a one-dimensional universe can split in two or two one-dimensional 
universes can merge to one. 
This is a kind of third quantization of 1+1 dimensional quantum gravity, 
and the corresponding string field theory has been formulated both 
for non-critical strings and GCDT \cite{kawai1,kawai2,kawai3,gcdt,cdtmatrix}. 
The emergent nature found in non-critical string theory 
is of Wilsonian nature: one starts with a lattice formulation 
which is in no way unique, but at certain critical points 
one can take continuum limits, 
such that  continuum theories of 2d quantum gravity coupled to 
conformal field theories will emerge. 
The lattice formulations of non-critical string theory 
were relatively closely related to underlying lattice
piecewise-linear geometries, and generalizations to 
higher-dimensional piecewise-linear geometries were suggested, 
leading to ``dynamical triangulated gravity''  (DT gravity) 
where the path integral over geometries was performed by summing over 
higher-dimensional triangulations constructed by gluing together 
higher-dimensional identical simplices. A modified version, generalizing 2d CDT 
to higher-dimensional triangulations also exists.
However, the ``emergence'' of  continuum theories of 
higher-dimensional quantum gravity at  critical points 
is still unclear (see \cite{ja} for a recent review of higher-dimensional DT and CDT). 
The so-called tensor models \cite{tensor} and group field theories \cite{GFT} 
represent further generalizations of  DT gravity, 
where the underlying graph structure is further away from 
any piecewise-linear geometry, but where one still hopes for 
the emergence of a higher-dimensional continuum quantum theories 
at critical points or at least continuum condensates \cite{condensate} (see \cite{recentcondensate}
for recent developments in this direction).\\
This brings us to another type of emergent structure. 
String theory, starting with the underlying strings, 
accommodates objects of various dimensions ($D$-branes), 
and there have been attempts to describe these using various matrix models, 
notably the BFSS  \cite{BFSS} and the IKKT matrix models \cite{IKKT}. 
These models are viewed as non-perturbative definitions of string theory 
and the ``emergence'' of (continuous) $D$-branes 
are then associated with certain sectors of these matrix models. 
Similarly, in the case of tensor models and group field theories, 
despite not defining any non-trivial continuum quantum field theories 
in the Wilsonian sense, as mentioned above certain sectors of these theories 
are typically promoted to continuum condensates, 
which are then given a cosmological interpretation. 
In this sense GCDT and the corresponding $W_3$-Jordan algebra theory 
presented here is slightly different and somewhat similar to the situation 
in string theory. 
The one-component theory is well defined as a continuum theory having 
a perturbative expansion in the genus of the $1+1$ dimensional universe. 
Moving to higher dimensions, one is led to multicomponent $W_3$ algebras
with Jordan algebra structure constants, as described below 
and in more detail in \cite{aw2}. 
The task is then to investigate if this multicomponent theory 
has any chance of describing our (higher-dimensional) universe, 
precisely the same task as is imposed on string theory. 
But contrary to string theory, once the theory is formulated as 
$W_3$-Jordan algebra theory, the natural starting point is not GCDT, 
but a purely pre-geometric algebraic theory with no time 
and no space interpretation. 
What we asked in \cite{aw2} was: how can such a pre-geometric theory 
``develop'' into a theory where time and space make sense, 
i.e.\ into a kind of multi-component GCDT theory, 
and next, how can such a multi-component theory develop 
into a theory that describes a higher-dimensional spacetime 
as we know it from our Universe? 
Crucial for this was the concept of ``knitting'' together 
the 1+1 dimensional universes, 
the dynamics of the knitting being dictated 
by the structure constants of the Jordan algebra.
}

The purpose of this article is two-fold. We want to understand better the 
knitting mechanism and we want to understand if the emergence  of 
the modified Friedmann equation, apart from fitting late time cosmology well,
 can also explain a number of  ``coincidences'' in our Universe. We will find that 
 the solution of the Friedmann equation is part of a algebraic curve where 
 a singular point has a special status that can be used to define the scale
 of our universe.

The rest of this article is organized as follows: in the next Section we shortly review 
the universe models based on Jordan algebras. 
As will be clear it is slightly misleading 
to talk about universe models, as they in fact are multiverse models. This is
how the wormholes enter. In Section \ref{knitting} we discuss possible 
outcomes of the knitting. 
In Section \ref{algebraic} we discuss how the modified Friedmann equation 
emerges and how the solution is part of an algebraic curve, the singular point 
of which defines a scale of our universe.
Sec.\ \ref{predictions} uses this scale to   explain 
the numerical coincidence of certain  numbers observed in cosmology. 
It further mentions how our model has the potential to explain 
the smallness of our coupling constant $g$ which replaces 
the cosmological constant of the standard $\L$CDM model.
{Further, we {\it suggest} a way} 
to explain the possible low entropy state of the early universe 
as well as the scale invariant fluctuations. 
Finally Sec.\ \ref{summary} summarizes our results.

\section{The Jordan algebra models of the Universe}\label{jordan}

The starting point of the Jordan algebra models is the two-dimensional 
so-called CDT model (CDT an abbreviation for Causal Dynamical Triangulations), 
\cite{al1}. The continuum limit of the triangulations  
can be viewed as a  version of 2d Horava-Lifshitz gravity \cite{agsw,horava}.
In particular it comes with a concept of proper time. 
It was generalized to allow for interaction of universes and 
creation and absorption of baby-universes, 
a model denoted GCDT (Generalized 
Causal Dynamical Triangulations) \cite{gcdt}, 
and this generalization could be formulated as a $W_3$ algebra  in 
much the same way as 
non-critical $c=0$ string theory \cite{mod1}. In \cite{aw1,aw2} we showed that the $W_3$ algebra encountered in GCDT 
can be extended to a  multicomponent $W_3$ algebra
provided the components form a Jordan algebra. The  number of components were, 
loosely speaking, associated with the number of spatial dimensions. 
More explicitly these higher 
dimensional GCDT models were defined by a Hamiltonian
\beq\label{jordan4}
{\hH} \propto
-\,\frac{1}{3} \sum_{k, l, m}  d_{abc}
:\!\alpha_k^{(a)} \alpha_l^{(b)} \alpha_m^{(c)}\!: \delta_{k+l+m,-2}.
\eeq
Here $d_{abc}$ denote the structure constants of the Jordan algebra with respect 
to some chosen orthonormal basis of the algebra (see \cite{aw2} for a detailed discussion).
$\alpha^{(a)}_k$denote standard creation and annihilation operators, i.e.
\beq\label{hj10} 
[\a_m^{(a)},\a_n^{(b)}] = m \,\del_{m+n,0} \,\delta_{a,b},
\eeq
and $: \cdots :$ denotes normal ordering of the $\a_m^{(a)}$. Summation over 
repeated indices $a,b,c$ in \rf{jordan4} is assumed.

The Jordan algebras that will have our main interest are the so-called spin factor type
algebras, related to the Clifford
algebras of $\g$-matrices, and algebras of Hermitian $3\times 3$ matrices 
where the entries belong to  $\Dbl{R}, \Dbl{C}, \Dbl{H}$ and $\Dbl{O}$, i.e.\ 
real,  complex,  quaternion and octonion  numbers, respectively.
The Clifford family of algebras 
are constructed from the real vector spaces 
generated by the $N$ traceless 
Hermitian  $\gamma$-matrices, $\g_a$, $a =1,\ldots, N$
of dimension $2^{[N/2]}$, and the identity matrix $\g_0$.

Ordinary matrix 
multiplication is defined for elements belonging to these vector spaces, but they become 
Jordan algebras with the following multiplication 
\beq\label{JordanProduct}
X \!\circ\! Y
\ \define \ 
\frac{1}{2} \{ X {\tinyspace},{\tinyspace} Y \}
\ \define \ 
\frac{1}{2} ( X Y + Y X )
\,.
\eeq
One can defined a real scalar product by 
\beq\label{jk22}
\la X ,Y \ra  \propto  \tr X\circ Y,
\eeq
and relative to this scalar product one can write 
\bea\label{jk24}
V_N &=&   \Dbl{R} \; I \oplus \tilde{V}_N,\label{jk26}\\
H_3(\Dbl{Q})  &=&  \versionII{\Dbl{R}}
  \; I \oplus \tH_3 (\Dbl{Q}), \label{jk25}
\eea
where the real vector space $\tilde{V}_N$ of traceless 
$\g$-matrices is $N$ dimensional and 
the real vector spaces of the traceless  
$3 \times 3$ Hermitean matrices $\tH_3 (\Dbl{Q})$ 
are 5, 8, 14 and 26 dimensional. 
The automorphism groups of the Jordan algebras are 
$\mathrm{SO}(N)$ and $\mathrm{SO}(3)$, $\mathrm{SU}(3)$, 
$\mathrm{Usp}(6)$ and $\mathrm{F}_4$, 
respectively.

In \cite{aw1,aw2} 
it was described how the Hamiltonian \rf{jordan4} could lead to 
a description of our present Universe. 
The line of argument is the following (we refer to \cite{aw1,aw2,w1} 
for details): viewing $\a^{(a)}_k$ as standard annihilation and 
creation operators defines a so-called absolute vacuum $| 0 \ra$. 
At this point $ {\hH}$
has no geometric interpretation. 
However, it is now assumed that the physical vacuum $\cuum$ 
is a coherent state, relative to the absolute vacuum, i.e.\ 
some of the annihilation operators $\a^{(a)}_k$ 
have  expectation values relative to the physical vacuum. 
In this way some of the cubic terms in $ {\hH}$ 
will appear as quadratic terms relative to the physical vacuum and 
$ {\hH}$ can be given an interpretation as an ordinary Hamiltonian 
with quadratic and cubic terms, except for the fact that 
${\hH}\cuum \neq 0$. More explicitly one can write 

\begin{equation}\label{DefAlphaOperatorCDT1}
\alpha_k^{(a)}
\,=\,
\expect{\alpha_k^{(a)}} + {\ha}_k^{(a)}
\,,
\end{equation}
and thus 
\begin{eqnarray}\label{WoperatorWithExpectationValue1}
\hH
{\negtrpltinyspace}\!\!&=&\!\!{\negdbltinyspace}
-g_0 \sum_{k,l,m} \delta_{k+l+m,-2}\, d_{abc} \left(
 \onethird  \expect{\alpha_k^{(a)}}
  \expect{\alpha_l^{(b)}}
  \expect{\alpha_m^{(c)}} 
+
 \expect{\alpha_{k}^{(a)}}
  \expect{\alpha_l^{(b)}}
  {\ha}_m^{(c)}\right.
\nonumber\\
&&\!\!{\negdbltinyspace}\phantom{%
-g_0 \sum_{k,l,m} \delta_{k+l+m,-2}\, d_{abc} \left(\right.
}\hspace{5pt}
+\left. \expect{\alpha_{k}^{a}}
  {\ha}_l^{(b)}{\trehalftinyspace}
  {\ha}_m^{(c)} 
+
\onethird   {\ha}_k^{(a)}{\trehalftinyspace}
  {\ha}_l^{(b)}{\trehalftinyspace}
 { \ha}_m^{(c)} \right)
\,.
\end{eqnarray}
The coupling constant $g_0$ of mass dimension 3 appears for dimensional reasons,
and the expectation values  $\la \a^{(a)}_k\ra$ can then be chosen such that 
some of the terms quadratic in  ${\ha}^{(a)}_k$ relate to 
standard kinetic terms,
while $g_0$ becomes a coupling constant multiplying the cubic 
${\ha}^{(a)}_k$ terms. Again, more explicitly, 
if $\la \a^{(a)}_{-3}\ra \neq 0$
for some values of the superscript $a$, 
the corresponding quadratic terms can be 
chosen to match the standard kinetic terms encounter in CDT. 
Similarly, some  $\la \a^{(a)}_{-1}\ra \neq 0$ can be chosen to match 
the cosmological constant terms encountered in CDT, either corresponding to
positive or negative cosmological constants. 
To be specific, the CDT Hamiltonian is 
\beq\label{e13}
\hH_{\rm CDT} = 
-\sum_{k=1}^\infty \ha_{k+1} \ha_{-k} + \mu 
 \sum_{k=2}^\infty \ha_{k-1} \ha_{-k}.
 \eeq 
Finally $\la \a^{(a)}_0\ra \neq 0$
will correspond to quadratic terms that are responsible of the 
creation or absorption of baby universes in GCDT. 
These terms will be important for the modified Friedmann equation 
to be discussed later 
(see \cite{mod1,mod2,mod3} for details) and in GCDT they have the form 
\beq\label{e14}
\versionII{\hH}_{\rm baby} = -2g_0 \sum_{k=3}^\infty \ha_{k-2} \ha_{-k}.
\eeq

The first term on the rhs of eq.~\rf{WoperatorWithExpectationValue1} 
is just a constant, while the terms linear in ${\ha}^{(a)}_k$ 
show that infinitesimal one dimensional spatial universes 
can be created from the physical vacuum. The reason for this is that 
some $\la \a^{(a)}_{-3}\ra$ and $\la \a^{(a)}_{-1}\ra$ are different from 
zero. Consequently some of the   ${\ha}^{(a)}_{k}$ operators
appearing among the linear terms have a subscript $k > 0$ and are thus
creating one-dimensional spatial universe of infinitesimal length. More precisely
the relations between   operators of a one-dimensional 
universe of macroscopic length $L$ and  operators $\ha_k$ are 
(see \cite{aw2}, eq. (2.49)) 
\beq\label{e11}
\Psi^\dg (L) = \sum_{k=0}^\infty \frac{L^k}{k!} \ha_k, \quad 
\ha_k = \frac{\dd^k}{\dd L^k} \Psi^\dg (L) \Big|_{L=0}, 
\quad 
L\Psi(L) = 
  \sum_{k=1}^\infty (-1)^{k}\ha_{-k} \Big(\frac{\dd}{\dd L}\Big)^{k} \del(L).  
\eeq 
The quadratic terms in the Hamiltonian can then propagate 
such an infinitesimal space in time to macroscopic size (see eq.\ \rf{e15} below).   
In this way the choice of 
a coherent state will first lead to the emergence of time (as the coefficient multiplying 
the Hamiltonian) and after that to the emergence of space. 
\versionII{
We should emphasize at this point that 
we unfortunately do not (yet) have a dynamical principle 
that will allow us to predict what kind of coherent state to choose. 
Clearly the emergent theory will depend on this choice. 
In \cite{aw2} some aspects of this are discussed in more detail.}
\versionII{
In addition, the time appearing as multiplying the Hamiltonian is identified with the global time that also appears in CDT and GCDT. This is natural as far as one considers
 a one-component model (i.e.\ essentially GCDT), but is strictly speaking an assumption when generalized to the multi-component Jordan algebra models. For a detailed discussion of this we refer to \cite{w1}.} 
For later reference
let us note that using \rf{e11} one can rewrite \rf{e13} and \rf{e14} in terms 
of the creation and annihilation operators of macroscopic loops
\beq\label{e11a}
\hH_{\rm CDT} = 
\int \frac{\dd L}{L} \Psi^{\dg}(L) 
\Big( -L \frac{\dd^2}{\dd L^2} + \mu L \Big)
L \Psi(L),
\eeq
\beq\label{e11b}
 \hH_{\rm baby} = 
-2g_0 \int \dd L  \;
\Psi^{\dg}(L)\Big( \frac{\dd}{\dd L}\Big)^{-1} \; L \Psi(L).
\eeq

If one, after the choice of physical vacuum, 
just looks at the quadratic Hamiltonian, 
one has a situation where the modes, 
labelled by the index $a$ from ${\ha}^{(a)}_k$, 
represent one-dimensional (closed) spatial universes of ``flavor'' $a$ 
propagating in time. 
These propagating 
spatial universes can either expand to infinite size 
in a finite time or stay of finite size, 
depending one the flavor $a$ and the choice of physical vacuum. 
This was described in some details in \cite{aw1,aw2,w1}, and let us here 
just mention that for the CDT Hamiltonian \rf{e11a} the time evolution 
$\la L(t) \ra$ for the spatial volume of a universe, 
starting at $t=0$ with $L(0) =0$ is
\beq\label{e15}
\la L(t) \ra = 
\frac{\tanh\!\big( \sqrt{\mu}\, t \big)}{\sqrt{\mu}}
 \quad {\rm or} \quad 
\la L(t) \ra = 
\frac{\tan\!\big( \sqrt{-\mu}\, t \big)}{\sqrt{-\mu}}
\eeq
for $\mu >0$ and $\mu < 0$, respectively.

However, such expanding spatial universes   
also interact due to the remaining cubic interaction terms.
In this way it is possible for two one-dimensional spatial universes of different flavors,
$a$ and $b$, to merge to a two-dimensional spatial universe labelled by $a,b$, and this
process can be generalized to $n$ different flavored one-dimensional universes 
merging to a $n$-dimensional spatial universe. This process was 
denoted {\it knitting} in \cite{aw1,aw2,w1}. We want to understand how this 
knitting leads to the state of our Universe. Since $ {\hH}\cuum \neq 0$ we have a 
non-trivial time evolution 
\beq\label{e1}
| \Psi(\T) \ra = \e^{- \T{\halftinyspace}{\hH}} \cuum
\eeq
that will describe the creation of microscopic one-dimensional universes and their
evolution, their growth and interaction. In principle we are thus dealing with a 
multiverse theory. Let us follow standard quantum field 
theory and introduce sources $j^{(a)}_k$ for these microscopic universes and the 
corresponding partition function
\beq\label{e2}
Z(j) =
 \vac  \; \e^{- \T{\halftinyspace}{\hH}}
 \e^{  \sum_{k> 0,a} {\ha}^{(a)}_k j^{(a)}_k} \cuum.
\eeq
The sources will provide us with a representation of the algebra \rf{hj10} in the 
standard way, where $j^{(a)}_n$ and $ \frac{\prt }{\prt j^{(a)}_n}$ 
act on suitable functions $F(j)$:
\bea
\a^{(a)}_n &\to& \frac{\prt }{\prt j^{(a)}_n} , \quad n > 0, \label{e3}\\
\a^{(a)}_n &\to& \la \a^{(a)}_n\ra -n  j_{-n} , \quad n < 0. \label{e4}
\eea
Using this representation the Hamiltonian \rf{WoperatorWithExpectationValue1}
will be a differential operator $\hH (j, \prt/ \prt j)$ 
acting on suitable functions $F(j)$.
As mentioned $ {\hH}\cuum \neq 0$, due to some of the linear ${\ha}_k^{(a)}$ terms
in \rf{WoperatorWithExpectationValue1}. 
As discussed in  \cite{aw1,aw2,w1}, if one
ignores these terms, one would have the ``no big-bang condition,''
\beq\label{e5}
\hH (j, \prt/ \prt j) Z(j) = 0.
\eeq
Ref.\ \cite{fmw12} further demonstrates that, 
under the no big-bang condition \rf{e5}, 
all CDT amplitudes can be obtained recursively.

Since the knitting is relating universes that have already been created, one 
might expect that if we only consider the knitting part of the Hamiltonian
one still has a relation like \rf{e5}, and we write tentatively
\beq\label{e6}
\hH_{\rm kn} (j, \prt/ \prt j) Z_{\rm kn}(j) = 0,
\eeq
where the subscript ``kn'' refers to ``knitting'' while
\beq\label{e7}
\hH_{\rm kn}({\ha}^{(a)}_k) 
= -\, \frac{g_0}{3} \sum_{k,l,m} \delta_{k+l+m,-2}\, d_{abc} 
  {\ha}_k^{(a)}{\trehalftinyspace}
  {\ha}_l^{(b)}{\trehalftinyspace}
 { \ha}_m^{(c)} 
\,.
\eeq
In eq.\ \rf{e6} we have used \rf{e3} and \rf{e4} 
to replace the $\ha$ operators in \rf{e7} 
with the operators $j$ and $\prt/\prt j$. 
As discussed above, while the $\ha$ operators 
act on the standard Fock space corresponding to 
an appropriate choice of vacuum state, 
the $j$ and $\prt/\prt j$ operators act on suitable functions $F(j)$. 
Thinking about the knitting as a dynamical process leads of course 
to an immensely complicated scenario, but if the end result 
is a higher dimensional spacetime one should probably think about this 
as a kind of time independent condensate, 
the time independence being expressed by eq.\ \rf{e6}.

We want to use eq.\ \rf{e6} to understand 
what the knitting implies for the function $Z_{\rm kn}(j)$.

\section{Knitting and exchange of flavors}\label{knitting}

\subsection{Spin factor algebras}

\begin{figure}
\centerline{ \includegraphics[width=200pt]{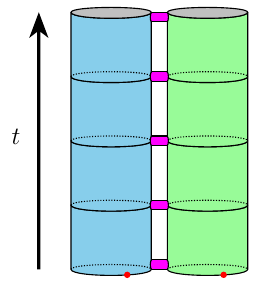}}
\caption{{\footnotesize
Points of two one-dimensional universes with different flavors (here different colors) are identified via knitting, i.e.\ by exchange of wormholes. In this way 
the two $T^1$ spatial universes merge into a a two-dimensional $T^2$ universe that 
propagates in time (the direction of the arrow). For combinatorial convenience
a point on the entrance loop of each of the  two universes  is marked 
(see \cite{al1,w1} for a discussion of this). 
}}
\label{fig1}
\end{figure}

For the purpose of illustration let us first consider the spin factor algebra case. Here the 
Jordan algebra structure constants can be chosen as 
\bea
&&
d_{000} \,=\, 1
\,,
\qquad
d_{0ab} \,=\, d_{a0b} \,=\, d_{ab0} \,=\, \delta_{ab}
\qquad\hbox{[\,$a$, $b {\negdbltinyspace}=\! 1$, \ldots, $N$\,]}
\,,
\nonumber\\
&&
\hbox{otherwise} \,=\, 0 \label{spinfactor}
\,.
\eea
If we choose 
\beq\label{e20}
\la \a^{(0)}_{-3} \ra= \frac{1}{2g_0}, 
\quad  
\la \a^{(0)}_{-1} \ra= - \frac{\mu}{2g_0},
\quad  
\la \a^{(0)}_{0} \ra= 1,
\eeq 
the quadratic term of each mode, for $a=0,1,\ldots,N$ will be like the 
CDT Hamiltonian \rf{e13}, and if $\mu > 0$ they will have a time evolution
as the one given in \rf{e15}, except that they will be interacting via the 
knitting Hamiltonian \rf{e7}, which in this case can be written as
\begin{equation}\label{ClifordNdimTypeModel1WoperatorInteraction}
\versionII{\hH}_{\rm kn} = -g_0 \Big(\frac{1}{3} \sum_{k=-\infty}^\infty \sum_{m=-\infty}^\infty
{\negqntpltinyspace}{\negqntpltinyspace}
 :
  \hat{\alpha}^{(0)}_{-2-k}
  \hat{\alpha}^{(0)}_{k-m}
  \hat{\alpha}^{(0)}_m
 :
\,+\,\,
2\!
\sum_{k=-\infty}^\infty
{\negqntpltinyspace}
  \hat{\alpha}^{(0)}_{-2-k} {\tinyspace}
  \hat{L}^{(1,\ldots,N)}_k 
\Big)  
\,,
\end{equation}
where the definition of $\hat{L}^{(1,\ldots,N)}_\ell$ is 
\begin{equation}\label{FlavorVirasoroOperatorInteraction}
\hat{L}^{(1,\ldots,N)}_\ell
\,\define\,
\half
\sum_{a=1}^N
\sum_{k=-\infty}^\infty
{\negqntpltinyspace}{\negqntpltinyspace}
 :
  \hat{\alpha}^{(a)}_{\ell-k}
  \hat{\alpha}^{(a)}_{k}
 :
\,.
\end{equation}
In this case the interaction picture is quite clear: the universes 
labelled by $a=1,\ldots,N$ can only interact via a ``wormhole'' 
with label 0, as shown in Fig. \ref{fig1}. 
Of course wormholes can also connect points on a spacetime with the 
same flavor, or two different spacetimes with the same flavor, but we consider 
that as being part of the standard ``gravity'' interaction, 
already present in GCDT, and not 
an interaction that will change the dimensionality of the given two-dimensional 
universe of a given flavor.

\begin{figure}
\vspace{-1.5cm}
\centerline{ \includegraphics[width=200pt]{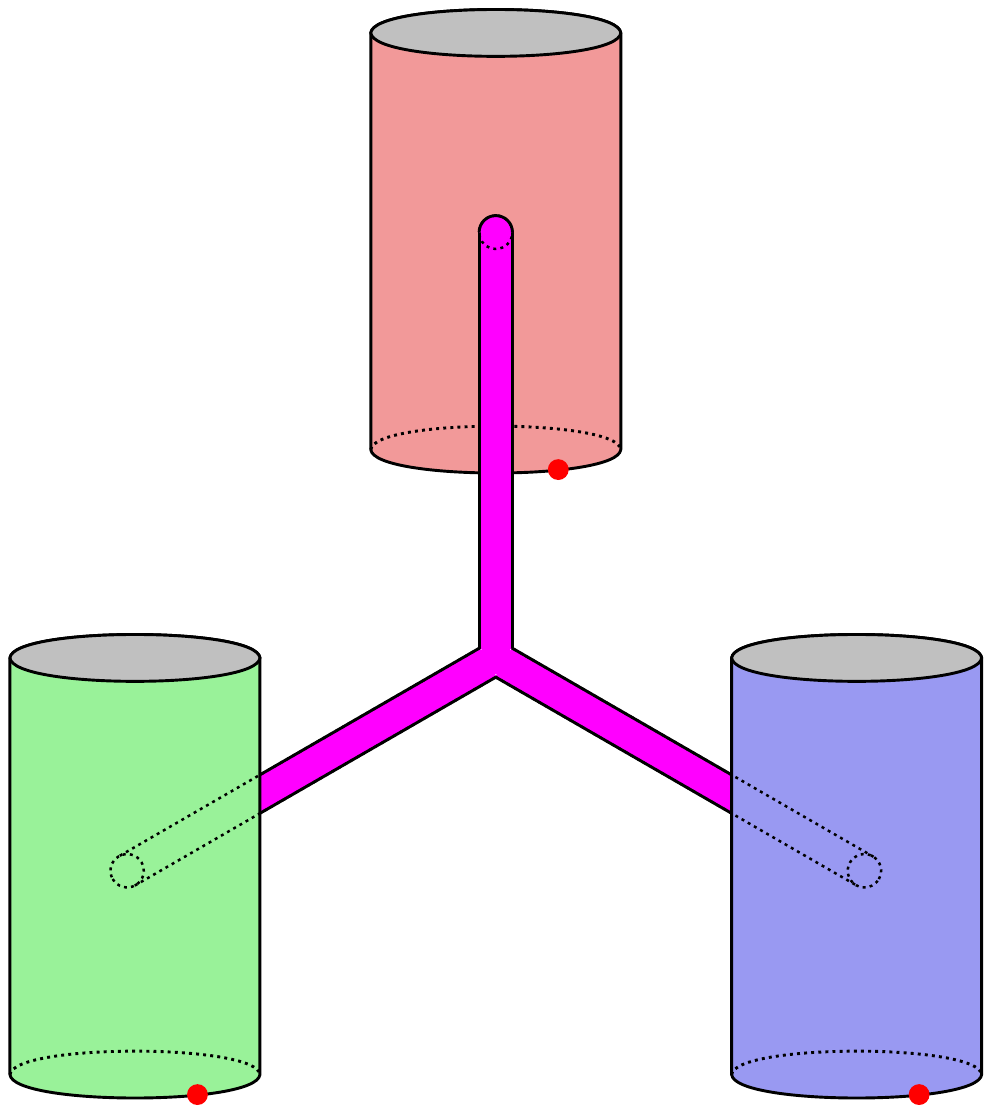}
\includegraphics[width=200pt]{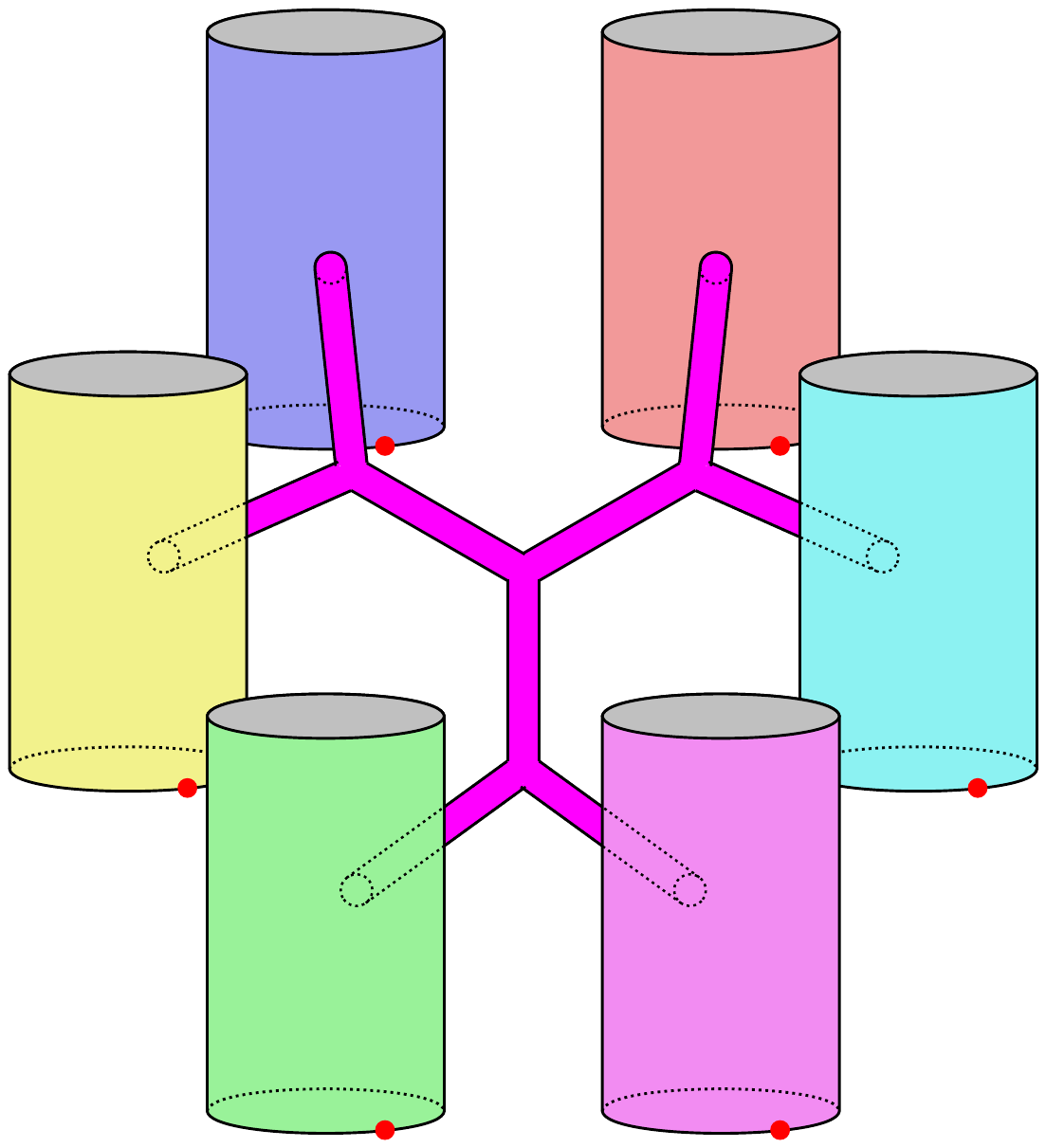}}
\vspace{-1.5cm}
\caption{{\footnotesize
On the left picture three one-dimensional universes with different flavors 
(here different colors) are identified via knitting, i.e.\ 
by exchange of wormholes. The wormholes forms a tree-diagram 
made possible by the cubic interaction term of the $\a^{(0)}$ mode 
in \rf{ClifordNdimTypeModel1WoperatorInteraction} . In this way 
three $T^1$ spatial universes merge into $T^3$, 
a  three-dimensional toroidal universe. 
The generalization to higher dimensional toroidal universes 
is shown on the right picture, 
where tree-diagrams of wormholes connect the universes of different flavors.
A $d$-dimensional universe is formed from tree diagrams with 
$N_{\rm edge} = d$ external vertices and thus 
$N_{\rm wh} = 2d-3$ wormholes,
interacting at $N_{\rm vertex} = d-2$ vertices.
}}
\label{fig2}
\end{figure}

Since the universes labelled 0 (the wormholes) have a cubic interaction, 
more complicated ``tree-diagrams'' can connect universes of various labels, 
as illustrated in Fig.\ \ref{fig2}. 
One can give qualitative arguments that 
tree-diagrams dominate over loop diagrams, 
but even these are not easily summed 
since the propagators corresponding to a Hamiltonian like \rf{e13} are 
quite complicated 
(see \cite{al1,gcdt} for expressions and simple calculations). 
However, if the end result of such a summation is 
a condensate where eq.\ \rf{e6} will be satisfied, 
this can be achieved by demanding 
\begin{equation}\label{ClifordNdimTypeModel1KNcond}
\hat{\alpha}^{(0)}_\ell {\tinyspace}Z_{\rm kn}({\tinyspace}j)= 0
\quad\hbox{[\,$\ell \ge 0$\,]}
\,,
\qquad\quad
\hat{L}^{(1,\ldots,N)}_\ell {\dbltinyspace}Z_{\rm kn}({\tinyspace}j) = 0
\quad\hbox{[\,$\ell \ge -1$\,]}
\,.
\end{equation}
The second condition is a kind of Virasoro-like constraint. 
Eq.\ \rf{ClifordNdimTypeModel1KNcond} allows a solution consistent 
with the $\mathrm{SO}(N)$ symmetry of the automorphism group. 
However, the choice of expectation values $\expect{\alpha_k^{(a)}}$ 
might break this symmetry 
and one should then choose different solutions to \rf{e6}. 
As a simple example one can in addition to \rf{e20} also choose 
\beq\label{e21}
  \la \a^{(N)}_{-1} \ra = - \frac{\mu_N}{2g_0},
\eeq
and the symmetry will be broken to $\mathrm{SO}(N\!-\! 1)$. 
Analyzing the quadratic terms as well as the knitting Hamiltonian as above 
then leads to a replacement of 
 \rf{ClifordNdimTypeModel1KNcond} by 
\beq
\hat{\alpha}^{(0)}_\ell {\tinyspace}Z_{\rm kn}({\tinyspace}j) = 
\hat{\alpha}^{(N)}_\ell {\tinyspace}Z_{\rm kn}({\tinyspace}j) = 0
\quad\!\!\hbox{[\,$\ell \ge 0$\,]} , \quad
\hat{L}^{(1,\ldots,N-1)}_\ell {\dbltinyspace}Z_{\rm kn}({\tinyspace}j) = 0
\quad\!\!\hbox{[\,$\ell \ge -1$\,]}. \label{e9}
\eeq
$\tau$-functions associated with Virasoro constraints like 
in eqs.\ \rf{ClifordNdimTypeModel1KNcond} or \rf{e9} are known 
from general Hermitian matrix models and the Kontsevich matrix model.
Since CDT can also be formulated as a matrix model \cite{cdtmatrix}, 
it is maybe not surprising 
that one encounters similar equations here, 
and it is an indication that  the $Z_{\rm kn}$ entering in these equations 
can be viewed as generalized $\tau$-functions, 
and thus that 
the stationary end result of the knitting is a condensate 
described by such a generalized $\tau$-function.
\versionII{
At this point we should emphasize that we have not derived dynamically that the knitting will result in a condensate where eq.\ \rf{e6} is satisfied, and while eqs.\ 
\rf{ClifordNdimTypeModel1KNcond} and \rf{e9} are sufficient conditions for 
eq.\  \rf{e6} to be satisfied, the corresponding generalized $\tau$-function will thus only represent a possible candidate condensate. Nevertheless we think it would be very interesting and potentially important to construct such a $\tau$-function explicitly.}

\subsection{The {$H_3(\Dbl{Q}) $} models}

 In the case of the Jordan algebra 
$H_3(\Dbl{Q})$, $\Dbl{Q} = \Dbl{R}$, $\Dbl{C}$, $\Dbl{H}$ or $\Dbl{O}$, 
one has a more complicated pattern for the structure constants $d_{abc}$. 
We refer to \cite{aw2,w1} for a detailed discussion. 
However, it is possible to make symmetry breakings 
that seem interesting from the point of view of cosmology 
and of string theory. 
Let us now describe such a pattern. 
We start by listing the $d_{\mu\nu \rho}$ structure constants 
in the case of $H_3(\Dbl{C})$, 
where they are just the Gell-Mann symmetric symbols 
if one chooses as $E_a$, $a =1,\ldots,8$ the Gell-Mann matrices $\l_a$ 
( and $E_0 = \frac{\sqrt{2}}{\sqrt{3}} I_{3\times 3}$):
\bea\label{e30}
  d_{000}\!\!\! &=&\! \!\!\sqrt{\frac{2}{3}},   \quad d_{0 ab} =
   \sqrt{\frac{2}{3}} \;\del_{ab}, ~
  a,b = 1,\ldots,8, \qquad d_{888} = -  \frac{1}{\sqrt{3}},\\
   \label{e31}
  d_{118}\!\!\!&=&\!\!\!d_{228} = d_{338} = \frac{1}{\sqrt{3}}, \quad
  d_{448}=d_{558} = d_{668}= d_{778} = -  \frac{1}{2\sqrt{3}},\\
  \label{e32}
  d_{443} \!\!\!&=\!\!\!&d_{553} = \oh, \quad d_{663}= d_{773} =- \oh, \\
  \label{e33}
  d_{146}\!\!\!&=&\!\!\!d_{157} = d_{256}=\oh, \quad  d_{247}= - \oh.
\eea
We now assume a symmetry breaking pattern similar to \rf{e20}:
\bea\label{e34}
 &&
 \la \a^{(0)}_{-3} \ra= \frac{\sqrt{3}}{\sqrt{2}}\,\frac{1}{2g_0}, 
 \quad  \la \a^{(0)}_{-1} \ra =  - \frac{\sqrt{3}}{\sqrt{2}}\;\frac{\mu_0}{2g_0},
 \quad  \la \a^{(0)}_{0} \ra =  \frac{\sqrt{3}}{\sqrt{2}},
\nonumber\\
 &&
 \la \a^{(8)}_{-1} \ra =  - \frac{\sqrt{3}\, \mu_8}{2g_0}, 
 \quad  \la \a^{(3)}_{-1} \ra =  - \frac{\mu_3}{2g_0}.
\eea
The symmetry breaking mass assignment in \rf{e34} is 
the most general one if the breaking takes place 
in a maximal Abelian subalgebra ($\l_3$ and $\l_8$ span 
a Cartan subalgebra of the Lie algebra $su(3)$ 
and any Cartan subalgebra can be obtained from it 
by a global $\mathrm{SU}(3)$ rotation, $\mathrm{SU}(3)$ 
being the automorphism group of  $H_3(\Dbl{C})$). 
As described in \cite{aw2} one can now choose values 
$\mu_0$, $\mu_8$ and $\mu_3$ such that diagonalizing the quadratic terms 
one obtains that the one-dimensional universes 
labelled 1 and 2 expand to infinity 
while the universes labelled 4, 5 as well as 6 ,7 remain of finite size.
In more detail,  when diagonalizing the quadratic terms 
involving the three modes labelled 0, 3 and 8 one obtains three modes
\beq\label{f1}
  \ha^{\versionII{([0])}}_k =
  \frac{1}{\sqrt{3}} \ha^{(0)}_k
  - \frac{\sqrt{2}}{\sqrt{3}} {\dbltinyspace} \ha^{(8)}_k,
  \quad
  \ha^{\versionII{([\pm])}} =
  \frac{1}{\sqrt{3}} \ha^{(0)}_k
  + \frac{1}{\sqrt{6}} \ha^{(8)}_k
  \pm \frac{1}{\sqrt{2}} \ha^{(3)}_k .
\eeq
The corresponding cosmological constants are 
\beq\label{f2}
  \mu_{[0]} = \mu_0 -2 \mu_8, \quad \mu_{[\pm]} = \mu_0 + \mu_8 \pm \mu_3,
\eeq
where the $[-]$ mode expands to infinity ($\mu_{[-]} < 0$) 
while the $[+]$ and the $[0]$ will only expand to finite size 
($\mu_{[0]},\; \mu_{[+]} > 0$) 
and act as wormholes knitting together the universes of different flavors, 
exactly like the 0-mode acted as the wormhole knitting 
together the rest of the one-dimensional universes 
in the spin factor model discussed above. 
For the 1 and 2 modes, the 4 and 5 modes and the 6 and 7 modes 
one obtains the cosmological constants
\beq\label{f3}
  \mu_{1,2} = \mu_0 + \mu_8,
  \quad
  \mu_{4,5} = \versionII{\mu_0 + \frac{-\mu_8 + \mu_3}{2}}
  ,
  \quad
  \mu_{6,7} = \versionII{\mu_0 + \frac{-\mu_8 - \mu_3}{2}}
  .
\eeq
For a suitable choice of $\mu_0,\mu_8,\mu_3$ one then obtains
\beq\label{f4a}
  \mu_{[0]} > \mu_{4,5} >
  \versionII{\mu_{6,7} > \mu_{[+]}}
  > 0 > \mu_{1,2} > \mu_{[-]}
\,,
\eeq
or
\beq\label{f4b}
  \mu_{[0]} > \mu_{4,5} >
  \versionII{\mu_{[+]} > \mu_{6,7}}
  > 0 > \mu_{1,2} > \mu_{[-]}
\,,
\eeq
\versionII{%
where $\mu_3 > 0$.
For $\mu_3<0$, one should interchange
$\mu_{[+]}$ with $\mu_{[-]}$ and $\mu_{4,5}$ with $\mu_{6,7}$.}
For the $H_3(\Dbl{C})$ model we thus end 
(including the time as one extended dimension) 
with a spacetime with four extended dimensions and 
6 compact dimensions. While plain numerology, 
one cannot avoid recalling the hope in superstring theory 
that its 10-dimensional spacetime will result 
in 4 extended dimensions and a six-dimensional compactified space, 
that is believed to be a Calabi-Yau manifold. 
In our knitting we argued that 
the knitted compact space should be a torus 
(see \cite{aw2,w1} for details), i.e.\ 
in the present case $\Dbl{T}^6$, which can be viewed as a Calabi-Yau manifold.

The above discussion for the Jordan algebra  $H_3(\Dbl{C})$ can be repeated 
for the other  $H_3(\Dbl{Q})$. The most interesting case is  $H_3(\Dbl{O})$,
the Jordan algebra based on octonions. Again, a discussed in \cite{aw2,w1},
the 0-8-3 system is unchanged while 
the universes with labels 1,2 are now extended to an octet with the 
negative cosmological constant $\mu_{1,2}$. Similarly, the 
universes with label 4,5 and with 6,7 are extended to two octets with two 
 positive cosmological constants $\mu_{4,5}$ and
$\mu_{6,7}$. In this way
we now obtain 9 spatial one-dimensional universes expanding to infinity and 
sixteen spatial dimensions staying compact, all knitted together by two wormhole 
modes. Again, using plain numerology, 
the 9+1 (time) extended directions and the 16 compact directions have some 
resemblance with the situation encounted in the bosonized version of the 
heterotic string model where one has a 10-dimensional extended spacetime and 
a compact 16-dimensional space with the topology of  $\Dbl{T}^{16}$.
\versionII{
It should be clear that, although tantalizing, this counting 
coincidence with string theory is presently nothing more than that. 
However, we are presently working on understanding 
if there can be a deeper connection.}

\subsubsection{Knitted states and Virasoro constraints}
 
\begin{figure}
\centerline{ \includegraphics[width=200pt]{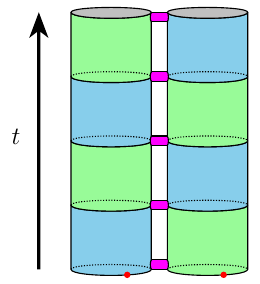}}
\caption{{\footnotesize
The figure is similar to Fig.\ \ref{fig1}, 
except that the interaction \rf{f9} will 
allow for the flavor to change during the propagation. 
Here we have only considered the oscillation 
between two flavors where the ``wormhole'' has a third flavor, 
but the structure constants $d_{iI\tilde{I}}$ 
allow for more complicated exchange patterns during the propagation.
}}
\label{fig3}
\end{figure}

We still want a relation like \rf{e6} to be satisfied, 
to ensure the stability of a knitting like the one described above 
for the spin factor model. 
Since the $W_3$ structure constants 
$d_{\mu\nu\rho}$ are more complicated for the $H_3(\Dbl{Q})$ models, 
also the interaction terms entering in
$\hH_{\rm kn}$ will be more complicated.  
After imposing the symmetry breaking 
\rf{e34}, the interaction consists of two parts, 
the first having the same structure 
as \rf{ClifordNdimTypeModel1WoperatorInteraction} only more elaborate, 
and we will
describe that now. 
Let us  denote 
the 1,2 modes by small latin indices $i$, 
the 4,5 modes by capital  indices $I$ and 
the 6,7 modes by capital indices $\tilde{I}$. 
In the case of octonions the 1,2 as well as 
the 4,5 and the 6,7 modes are extended to octets. 
The knitting Hamiltonian \rf{e7} generalizes now to 
 \bea\label{f5}
 \hH_{\rm kn} &=&
 -g_0 \left( \frac{1}{3}
 \sum_{k=-\infty}^\infty \sum_{m=-\infty}^\infty d_{\a\b\g}
 : \hat{\alpha}^{(\a)}_{-2-k}
  \hat{\alpha}^{(\b)}_{k-m}
  \hat{\alpha}^{(\g)}_m: + 
  \sum_{k=-\infty}^\infty
  \big(\ha_{-2-k}^{([+])} +\ha_{-2-k}^{([-])}\big)
  \hat{L}^{1,2}_k \right.
\nonumber\\
   &&\phantom{%
 -g_0 \left(\right.
}%
+
   \left. \frac{2}{3}\sum_{k=-\infty}^\infty \big(\ha_{-2-k}^{([0])} +\ha_{-2-k}^{([+])}
  - \oh  \ha_{-2-k}^{([-])}\big) \hat{L}^{4,5}_k  \right.
\nonumber\\
   &&\phantom{%
 -g_0 \left(\right.
}%
+
   \left. \frac{2}{3} \sum_{k=-\infty}^\infty  \big(\ha_{-2-k}^{([0])} +\ha_{-2-k}^{([-])}
  - \oh  \ha_{-2-k}^{([+])}\big)
  \hat{L}^{6,7}_k \right) ,
  \eea
  where the $(\ha^{(\a)}\ha^{(\b)}\ha^{(\g)}$ term is the sum over the  cubic terms, 
  with the $\ha^{(\a)}$, $\a = [0],[\pm]$ defined in eq.\ \rf{f1}, and the corresponding 
  $d_{\a\b\g}$ are only different from zero if $\a=\b=\g$.
  The $\versionII{\hat{L}}$ terms 
  are defined in analogue with \rf{FlavorVirasoroOperatorInteraction}
\beq\label{f6}
  \versionII{\hat{L}}^{1,2}_k =
  \oh \sum_{l=-\infty}^{\infty} :\ha^{(i)}_{k-l} \ha^{(i)}_l\!:, 
\quad
  \versionII{\hat{L}}^{4,5}_k =
  \oh \sum_{l=-\infty}^{\infty} :\ha^{(I)}_{k-l} \ha^{(I)}_l\!:, 
\quad 
  \versionII{\hat{L}}^{6,7}_k =
  \oh \sum_{l=-\infty}^{\infty} :\ha^{(\tilde{I})}_{k-l} \ha^{(\tilde{I})}_l\!: .
\eeq
In addition to the sum over $l$, one should sum over $i$, $I$ and $\tilde{I}$,
which in the case of $H_3(\Dbl{C})$ are doublets and for $H_3(\Dbl{O})$
octets. Eq.\ \rf{e6} now leads to the equivalent to 
\rf{ClifordNdimTypeModel1KNcond}, but for the $H_3(\Dbl{C})$ and the 
$H_3(\Dbl{O})$ models:
\beq\label{f7}
\hat{\alpha}^{([0])}_\ell {\tinyspace}Z_{\rm kn}({\tinyspace}j)=
\hat{\alpha}^{([+])}_\ell {\tinyspace}Z_{\rm kn}({\tinyspace}j)=
\hat{\alpha}^{([-])}_\ell {\tinyspace}Z_{\rm kn}({\tinyspace}j)= 0
\quad\hbox{[\,$\ell \ge 0$\,]}, 
\eeq
\beq\label{f8}
\hat{L}^{1,2}_\ell {\dbltinyspace}Z_{\rm kn}({\tinyspace}j) = 
\hat{L}^{4,5}_\ell {\dbltinyspace}Z_{\rm kn}({\tinyspace}j) =
\hat{L}^{6,7}_\ell {\dbltinyspace}Z_{\rm kn}({\tinyspace}j)= 0
\quad\hbox{[\,$\ell \ge -1$\,]}
\,.
\end{equation}
While this $\hH_{\rm kn}$ from \rf{f5} is 
the direct analogue of the ``knitting Hamiltonian'' 
of the spin factor model, coming from structure constants
$d_{\mu \nu \rho}$, we also have a ``residual'' interacting coming from 
structure constants where all indices are different.  
These structure constants and the corresponding interactions have the form
\beq\label{f9}
\hH_{\rm ex} = 
-\,2 g_0 \sum_{i,I, \tilde{I}} d_{iI\tilde{I}}\sum_{k=-\infty}^{\infty} 
\sum_{m= -\infty}^{\infty}
 :\ha_{-2-k}^{(i)} \ha^{(I)}_{m-k} \ha^{(\tilde{I})}_m\!:.
\eeq
We denote this part of the interaction the ``exchange Hamiltonian", 
since it potentially can exchange propagating modes of type $i$ 
with modes of type $I$ or $\tilde{I}$.
This is illustrated in Fig.\ \ref{fig3}. 
If spatial distances are very small and $g_0 \neq 0$
it thus make no sense to talk about individual modes of 
type $i$, $I$ and $\tilde{I}$
since they can mix while propagating. 
However, the modes of type $i$ can expand to macroscopic size 
(as will happen if the coupling constant $g_0$ is small), and 
once this has happened $ \hH_{\rm ex}$ can be ignored. 
The reason for this is that 
the spatial volume $L$ is conserved in the interaction 
(for a detailed discussion see \cite{aw2,w1}, 
where the interaction term is written in terms of 
the fields \rf{e11} that depend on $L$). 
Thus, as an example, 
a universe of type $i$ will only very rarely be able to ``decay'' 
into two  universe of type $I$ and $\tilde{I}$, 
since then (at a given time)
\beq\label{f10}
  \la L(i) \ra \gg \la L(I) \ra, \la L( \tilde{I}) \ra\quad {\rm and}\quad
  L(i) = L(I) + L(\tilde{I}).
\eeq
To summarize: 
if extended universes have managed to form, the knitting mechanism 
might work well and \rf{f7} and \rf{f8} be satisfied, while  $\hH_{\rm ex}$
will not be important. However, at very early times a complicated dynamics 
might prevail that we will now try to address.

\subsubsection{Exchange states and Virasoro constraints}

We will show that under certain assumptions \rf{f9} can be simplified and 
recasted into a form involving Virasoro operators, 
much like the ones that appeared in the knitting Hamiltonian. 
The assumptions are, loosely speaking, that some of the flavors 
can be identified. 
Again we will use $H_3(\Dbl{C})$ as an example, but 
the arguments will be valid for the other $\Dbl{Q}$ models with minor changes. 

Similar to \rf{e6} and \rf{e7} we write
\beq\label{h1}
  \hH_{\rm ex} (j, \prt/ \prt j) Z_{\rm ex}(j) = 0,
\eeq
where $\hH_{\rm ex}$ is defined by \rf{f9} with 
$\a \to j,~\partial/\partial j$ (as in eqs.\ \rf{e3} 
and \rf{e4}).
A first case to discuss is 
when we partly identify flavors with index $I$ with flavors with index 
$\tilde{I}$, more specifically 4 with 6 and 5 with 7. 
In order to satisfy \rf{h1} we assume
\beq\label{h2}
\ha^{(4)}_l Z_{\rm ex}(j) = \ha^{(6)}_l Z_{\rm ex}(j), \quad 
\ha^{(5)}_l Z_{\rm ex}(j) = \ha^{(7)}_l Z_{\rm ex}(j),
\eeq
\beq\label{h3}
\ha^{(1)}_l Z_{\rm ex}(j) =0,
\quad l \geq 0.
\eeq
We write
\bea\label{h4}
  \hH_{\rm ex} &\propto& \sum_{k=-\infty}^{\infty} 
  \sum_{m= - \infty}^{\infty}
    \ha^{(1)}_{-2-k}\big(\ha^{(4)}_{m-k} \ha^{({6})}_m + \!\ha^{(5)}_{m-k}
    \ha^{({7})}_m\big) +\\
  &&\sum_{k=-\infty}^{\infty} 
  \sum_{m= - \infty}^{\infty} \!
    \ha^{(2)}_{-2-k}(-\ha^{(4)}_{m-k}
    \ha^{({7})}_m \!+\! \ha^{(5)}_{m-k} \ha^{({6})}_m).
  \nonumber
\eea
Eqs.\ \rf{h2} and \rf{h3} allow us to write 
\beq\label{h5}
  \hH_{\rm ex} Z_{\rm ex} \propto 
\sum_{k=-\infty}^{\infty}  \ha^{(1)}_{-2-k} \; \versionII{\hat{L}}_k^{4,5} Z_{\rm ex} ,
\eeq
and we can obtain 
\beq\label{h6}
  \hH_{\rm ex} Z_{\rm ex} = 0
  \quad
  {\rm if} \quad  \versionII{\hat{L}}_k^{4,5} Z_{\rm ex} = 0
  \quad
  {\rm for} \quad k \geq -1,
\eeq
in addition to \rf{h2} and \rf{h3}, i.e.\ 
if Virasoro-like constraints are satisfied.

An similar construction can be obtained by 
identifying flavors of index $i$ with flavors of index $I$. 
Let us identify flavor 1 with flavor 4 and flavor 2 with flavor 5.
We now impose  
\beq\label{h7}
  \ha^{(1)}_l Z_{\rm ex}(j) = \ha^{(4)}_l Z_{\rm ex}(j),
  \quad 
  \ha^{(2)}_l Z_{\rm ex}(j) = \ha^{(5)}_l Z_{\rm ex}(j),
\eeq
\beq\label{h8}
  \ha^{(6)}_l Z_{\rm ex}(j) =0,
  \quad l \geq 0,
\eeq
and obtain 
\beq\label{h9}
  \hH_{\rm ex} Z_{\rm ex} 
  \propto 
  \sum_{k=-\infty}^{\infty}  \ha^{(6)}_{-2-k} \; \versionII{\hat{L}}_k^{1,2} Z_{\rm ex}
  =
  \sum_{k=-\infty}^{\infty}  \ha^{(6)}_{-2-k} \; \versionII{\hat{L}}_k^{4,5}Z_{\rm ex} ,
\eeq
\beq\label{h10}
  \hH_{\rm ex} Z_{\rm ex} = 0
  \quad
  {\rm if} \quad
  \versionII{\hat{L}}_k^{1,2} Z_{\rm ex} = 0
  \quad
  {\rm for} \quad k \geq -1,
\eeq
i.e.\ again a Virasoro-like constraint. 
Similarly, we can identify flavors $i$ with flavors $\tilde{I}$, 
for instance  flavor 1 and 7 as well as 2 and 6, 
in which case we  obtain the analogue of \rf{h7}-\rf{h10}:
\beq\label{h11}
  \ha^{(1)}_l Z_{\rm ex}(j) = \ha^{(7)}_l Z_{\rm ex}(j),
  \quad 
  \ha^{(2)}_l Z_{\rm ex}(j) = \ha^{(6)}_l Z_{\rm ex}(j),
\eeq
\beq\label{h12}
  \ha^{(5)}_l Z_{\rm ex}(j) = 0,
  \quad
  l \geq 0,
\eeq
\beq\label{h13}
  \hH_{\rm ex} Z_{\rm ex} \propto 
  \sum_{k=-\infty}^{\infty}
    \ha^{(5)}_{-2-k} \; \versionII{\hat{L}}_k^{1,2} Z_{\rm ex} =
  \sum_{k=-\infty}^{\infty}
    \ha^{(5)}_{-2-k} \; \versionII{\hat{L}}_k^{6,7}Z_{\rm ex} ,
\eeq
\beq\label{h14}
  \hH_{\rm ex} Z_{\rm ex} = 0
  \quad
  {\rm if} \quad  \versionII{\hat{L}}_k^{1,2} Z_{\rm ex} = 0
  \quad
  {\rm for} \quad k \geq -1.
\eeq

One difference between the scenario where flavors of index $I$ are 
identified with flavors of index $\tilde{I}$ 
and the scenarios where flavors of index $i$ are 
identified with either flavors of index $I$ or index $\tilde{I}$ is that 
the first scenario is compatible with the evolution of flavored universes 
hinted by the kinetic term: the universes of index $I$ and $\tilde{I}$ 
stay bounded in the time evolution induced by their kinetic terms. 
In the other scenarios universes with index $i$ 
will expand to infinite size (in a finite time),
contrary to a universe with indices $I$ or $\tilde{I}$. 
Thus an identification of flavors with $i$ index with flavors 
with $I$ or $\tilde{I}$ indices is only natural 
if we can ignore the kinetic term. 
If time is sufficiently short one can probably do that, 
and it is then possible that an expanding $i$ universe 
can be locked to a non-expanding $I$ or $\tilde{I}$ universe. 
This would be the case if the two universes were created 
at different times where the time difference is so small that 
$\hH_{\rm ex}$ from \rf{h9} or \rf{h13} would dominate the kinetic term.
However, if universes with flavor $i$ have time to grow sufficiently, 
we end up in a situation like in \rf{f10} and then $\versionII{\hH}_{\rm ex}$ 
will probably play no role, as discussed above. 
\rm Before, when we discussed the knitting mechanism of universes 
with different flavors, 
we had in the case of $H_3(\Dbl{C})$ an extended spacetime of 4 
and the rest of the dimensions were compact. 
But by the above mechanism 
either one or two of extended 1,2 flavors can remain compact, 
leaving us with 2 or 3 dimensional extended spacetime. 
In the case of $H_3(\Dbl{O})$, 
the 1,2-flavors are extended to an octet, 
so the starting point was a 10 dimensional spacetime, 
but by locking any number of the 8 extended flavor dimensions 
to the compact flavors we can end up with an extended spacetime 
being anything from 3 to 10. 
Unfortunately, it is clear that the above arguments are in no way rigorous, 
but they hint that both the $H_3(\Dbl{C})$ and the $H_3(\Dbl{O})$ models 
might be able to generate universes with 4 extended spacetime dimensions.

\subsubsection{Averaging of cosmological constants by the exchange mechanism}

For two spaces with flavors A and B
that form a pair via the exchange mechanism,
the identification
$\versionII{\hat{\alpha}}_n^{(\mathrm{A})} \sim
 \versionII{\hat{\alpha}}_n^{(\mathrm{B})}$
allows the operator corresponding to the cosmological constant
to be rewritten as
\begin{eqnarray}
&&
    \mu^{({\rm A})} \sum_{n=1}^\infty
    \versionII{\hat{\alpha}}_{-n-1}^{({\rm A})}
    \versionII{\hat{\alpha}}_n^{({\rm A})}
  + \mu^{({\rm B})} \sum_{n=1}^\infty
    \versionII{\hat{\alpha}}_{-n-1}^{({\rm B})}
    \versionII{\hat{\alpha}}_n^{({\rm B})}
\nonumber\\
&&\sim\ 
    \frac{\mu^{({\rm A})}+\mu^{({\rm B})}}{2}
    \sum_{n=1}^\infty
      \frac{\versionII{\hat{\alpha}}_{-n-1}^{({\rm A})} +
            \versionII{\hat{\alpha}}_{-n-1}^{({\rm B})}}{\sqrt{2}}
      \frac{\versionII{\hat{\alpha}}_n^{({\rm A})} +
            \versionII{\hat{\alpha}}_n^{({\rm B})}}{\sqrt{2}}
\,,
\end{eqnarray}
so that the cosmological constant is effectively given by their average value.

In particular, in the $0$--$8$--$3$ system of $H_3(\Dbl{O})$,
one considers two sets of one-dimensional spaces:
the set of eight flavors ($i$) with negative cosmological constants
$\mu_{1,2}<0$,
and the set of sixteen flavors ($I,\tilde{I}$)
with positive cosmological constants
$\mu_{4,5}=\mu_{6,7}>0$.
The average of the cosmological constants
associated with these two flavor sets,
as induced by the exchange mechanism,
is given by
\[
( \mu_{1,2} + \mu_{4,5} ) / 2
= ( \mu_{1,2} + \mu_{6,7} ) / 2 > 0 .
\]
Accordingly, when a space from the $i$-sector forms a pair with a space
from the $I$ or $\tilde{I}$ sector via the exchange mechanism,
the effective cosmological constant becomes positive,
and the expansion of the paired spaces settles into a tanh-type behaviour.

However, the exchange mechanism mediated by only a single wormhole 
is intrinsically much weaker than the knitting mechanism. 
Moreover, the effective coupling constant $g$ that connects 
spaces with different flavors through the knitting mechanism 
(See subsection \ref{sec:smallg}.)~takes an effectively small value. 
Consequently, once a space has grown far beyond the Planck scale, 
pair formation via the exchange mechanism is strongly suppressed, 
and the subsequent time evolution is expected to be dominated 
by expanding solutions governed by classical gravitational equations.

An important point to note here is that
each one-dimensional space is created as an individual event,
with a random creation time that does not depend on its flavor.
As a consequence,
a temporal mismatch can arise between the creation of spaces
belonging to the eight-flavor set ($i$) with negative cosmological constant
and those belonging to the sixteen-flavor set ($I,\tilde{I}$)
with positive cosmological constant.

During stages in which the corresponding partner spaces
have not yet been created,
spaces with negative cosmological constant are not stabilized
by the exchange mechanism
and can therefore grow beyond the Planck scale.
As a result, a subset of the flavor-$i$ spaces
can expand to macroscopic size.
On the other hand, the flavor-(3) space does not participate
in pair formation via the exchange mechanism and hence
continues to grow beyond the Planck scale without obstruction.

Taken together, these considerations imply that
universes whose spatial dimensionality ranges from one to nine
can emerge as classically growing configurations.
Moreover, the probability for the emergence of spaces
with each dimensionality becomes theoretically computable
within the present framework.
%
%
%
%
To estimate the resulting probabilities, we introduce a simplifying assumption for the sake of approximation: the one-dimensional spaces carrying the 24 flavors are generated independently, one for each flavor, and with equal probability\footnote{ 
\versionII{
The assumption that the three octets are created with equal probability 
should be regarded as a zeroth-order approximation, 
and it is not without justification.
Although the universe emerges from nothing, 
the octets are not created directly. 
First, the three singlets are generated from the vacuum through the terms in 
\beq
  \frac{1}{\sqrt{2}}
   \Big\{\!
-\frac{1}{4 g_0} \big(
        \hat{\alpha}^{([0])}_{4}
        + \hat{\alpha}^{([+])}_{4} + \hat{\alpha}^{([-])}_{4}
    \big)
    - \big(
        \hat{\alpha}^{([0])}_{1}
        + \hat{\alpha}^{([+])}_{1} + \hat{\alpha}^{([-])}_{1}
    \big)
      + \frac{1}{2 g_0} \big(
        \mu_{[0]} \hat{\alpha}^{([0])}_{2}
      + \mu_{[+]} \hat{\alpha}^{([+])}_{2}
      + \mu_{[-]} \hat{\alpha}^{([-])}_{2}
    \big)
   \!\Big\}
\nonumber
\eeq
}
\versionII{
The first terms are identical for the three singlets while the last terms differ because 
the cosmological constants can differ.
Subsequently, octets are pair-created from the three singlets through
\begin{align}
- \sqrt{2}\;\! g_0 \sum_k \sum_m \Big\{
&
    \sum_i
      \hat{\alpha}^{(i)}_{-2-k} \hat{\alpha}^{(i)}_{k-m}
      \big( \hat{\alpha}^{([+])}_m + \hat{\alpha}^{([-])}_m \big)
   +
    \sum_I
      \hat{\alpha}^{(I)}_{-2-k} \hat{\alpha}^{(I)}_{k-m}
      \big( \hat{\alpha}^{([0])}_m + \hat{\alpha}^{([+])}_m \big)
\nonumber\\
&
   +
    \sum_{\tilde{I}}
      \hat{\alpha}^{(\tilde{I})}_{-2-k} \hat{\alpha}^{(\tilde{I})}_{k-m}
      \big( \hat{\alpha}^{([0])}_m + \hat{\alpha}^{([-])}_m \big)
  \Big\}
\,.\nonumber
\end{align}
Therefore, the creation probabilities of the three octets 
are expected to differ slightly because of the differences 
between the cosmological constants $\mu_{[0]}$, $\mu_{[+]}$ and $\mu_{[-]}$.
\\
In this way, the three octets are not produced in a single interaction. 
An exact calculation of their creation probabilities 
is therefore not straightforward. 
However, as long as the values of the cosmological constants 
are of the same order of magnitude, 
the differences in the creation probabilities are expected to be 
relatively small. 
We therefore adopted equal probabilities as a zeroth-order approximation.
In addition to this assumption, we assumed that 
the probability is unity that an earlier-born flavor universe, 
which has not yet been paired, captures a later-born flavor universe 
and pairs with it.
}%
} 
Under this approximation, 
the probabilities for the emergence of universes with one-, two-, and three-dimensional spaces are estimated to be approximately $0.529$, $0.288$, and $0.125$, respectively. For higher-dimensional spaces the probability is further suppressed, and in particular the realization of a universe with a nine-dimensional space becomes exceedingly unlikely.

Although this probability distribution may appear to lack physical meaning
from the viewpoint that we can only observe our own Universe
with three spatial dimensions,
this is not the case.
Through the Coleman mechanism,
our Universe interacts---via long and thin wormholes---with 
many spaces of different dimensionalities.
As a result, the probability distribution for the generation of spaces
with various dimensions influences the dynamics
at the end of inflation.
In this sense, the distribution can be indirectly reflected
in observable quantities.

In the following 
we will assume that the dimensionality of the extended spacetime 
is four, and will discuss the simplest such situation 
where this spacetime has maximal symmetry.

\section{The Universe as an algebraic curve}\label{algebraic}

Let us now now return to \rf{e14} and its expression \rf{e11b} in terms of macroscopic
loops. We denote this  Hamiltonian, 
the quadratic term arising from assigning an expectation value to 
$\a_0$, the baby universe Hamiltonian. The reason for the name is that it can  be 
viewed as the limit $L \to 0$ limit of the non-local quadratic term appearing 
when one assigns an expectation value $\la L \Psi(L) \ra = \a(L)$ to 
$L \Psi(L)$, i.e.\ assuming $\a(L) = \a_0 + \a_1L + a_2 L^2 + \cdots$, 
as described in detail in \cite{mod3}.  
If we identify $\dd/\dd L = P$, eqs.\ \rf{e11a} and \rf{e11b} allow us 
(after a suitable rotation to Minkowskian signature, described in detail 
in \cite{mod3}) to identify a classical Hamiltonian 
\beq\label{g1}
 H = L\Big(\!\! -P^2 + \mu_0 - \frac{2g_0}{P} \Big).
\eeq  
This Hamiltonian is derived for GCDT, 
i.e.\ a theory of two-dimensional spacetime
with no $W_3$ flavor indices\footnote{As described in \cite{mod3}, assuming 
an expansion $\la L \Psi(L) \ra = \a(L)= \a_0 + \a_1 L + \a_2 L^2+ \cdots$ 
leads to a power series in inverse powers of $P$ in \rf{g1}, 
starting with the term $1/P$ associated to $\a_0$.}. 
Once one considers the $W_3$ theory with several flavors, 
we have different $L^{(a)}$, different momenta $P^{(a)}$ and 
different cosmological constants, 
and as we have seen there can be several outcomes of the knitting 
and the exchange mechanism.
As mentioned  we will here assume that the outcome 
is the appearance of a macroscopic space of dimension 3.
\versionII{Further we assume that after knitting there will be a mechanism like Coleman's
\cite{coleman} that will be active for macroscopic spaces, i.e. in our context after knitting has taken place. It is  an added assumption which is presently 
unrelated to our $W_3$/Jordan Algebra model.}
 This implies that 
effectively their cosmological constants, 
while starting out being different from zero 
and even different for certain space directions, 
in the end can be viewed as being zero.
\versionII{
Assuming maximal symmetry of the 1+3 dimensional space time 
formed by knitting, i.e.\ that 
the metric can be represented by the FLRW metric, 
it is natural to assume that 
the 1+1 dimensional Hamiltonian (60) survives in the form 
where the spatial length $L$ is replaced by the spatial volume $V$. 
This leads to eq.\ (61) below. 
Apart from being a generalization of (60) to $d > 1$ spatial dimensions, 
the kinetic term proportional to $V P^2$ is at the same time also 
the Hartle-Hawking minisuperspace Hamiltonian of GR for $d>1$ 
in the FLRW metric,}
\beq\label{g2}
  H_{\rm eff} =
  V \Big(\! - \frac{3}{4}\big( P^2 +  \frac{2g}{P}\big) \Big),
  \quad
  V = \frac{1}{\k_3} L^3, \quad \k_3 = 8\pi G_{\rm N},
\eeq
where $L$ is the scale factor in the FLRW metric 
and $V$ denotes the corresponding ``volume'' factor. 
$G_{\rm N}$ is the gravitational  constant and 
we have divided $V$ by $\k_3$ for convenience, 
to limit the appearance of $G_{\rm N}$ 
in some of the formulas to come. $P$ is the conjugate momentum to $V$. 
\versionII{The natural generalization of \rf{g1} to higher dimensions would in addition 
to the terms in \rf{g2}
also include a cosmological term proportional to $V\,\Lambda$. However, as mentioned above, 
after the knitting is completed (when \rf{g2} should be valid), we assume that Coleman's mechanism
effectively puts such a cosmological constant $\Lambda =0$.}
The factor $3/4$ is just convention, 
such that the $P^2$ term agrees with 
the Hartle-Hawking minisuperspace Hamiltonian in four-dimensional spacetime.

An important point is that the $g$ appearing in \rf{g2} 
is not the same as the $g_0$ appearing in \rf{g1}. 
The reason is that the Hamiltonian appearing in \rf{g2} is the one 
obtained by knitting and in this process $g_0$ can be ``renormalized''. 
We will discuss this in some detail in Sec.\ \ref{predictions}. 
We will think of $g_0$ as being of order 1 while $g$ can be very small.

As discussed in \cite{mod3,w1} the Hamiltonian \rf{g2} allows us 
to maintain the interpretation of \rf{g1} as a Hamiltonian 
where baby universes of infinitesimal length $\del L$ are being absorbed 
by the ``parent'' universe of length $L$, only with the length $L$ 
being replaced by the volume $V$ of the higher dimensional space.  
Also, as discussed in \cite{mod3,w1}, despite 
the missing cosmological term, the last term in \rf{g2} will result 
in a late time exponentially expanding universe, the expansion caused 
by the absorption of baby universes. 

\begin{figure}
\centerline{\includegraphics[width=300pt]{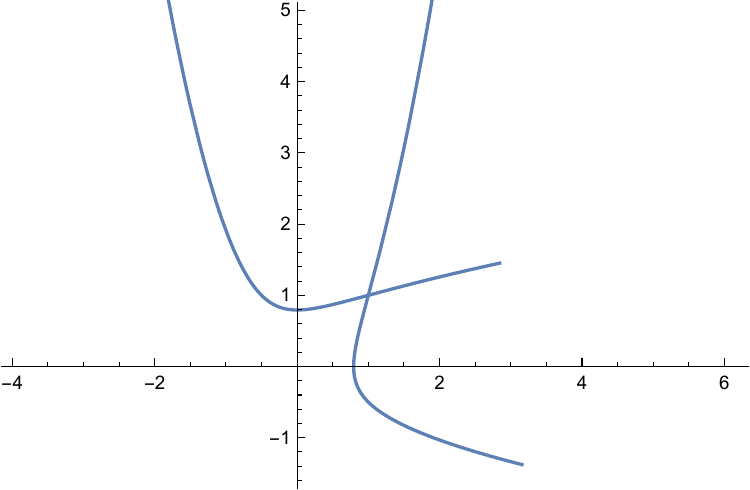}
}
\caption{{\footnotesize
The algebraic curve \rf{g8} plotted in the $(\l,h)$ plane. 
The branch of the curve parametrized by negative $p$ starts 
at the left at $(\l,h) \to (-\infty,\infty)$  for $p \to -0$ 
and move to the right for decreasing $p$.  
The branch of the curve parametrized by positive $p$ 
goes to $(\infty,\infty)$ for $p \to +0$, and to $(\infty, -\infty)$ 
for $p \to \infty$. 
The two branches intersect in (1,1).
}}
\label{fig4}
\end{figure}

It is now possible to consider the late time cosmology 
created by $H_{\rm eff}$, 
provided we add cold dark matter (CDM) to the Hamiltonian 
and compare the prediction of that model to 
the standard $\L$CDM model predictions.
Surprisingly, it is doing better than the $\L$CDM model {\it provided} one 
assumes that the Hubble constant  $H_0$ is 
73 \versionII{km/s/Mpc}\footnote{If one does not insist that 
the locally measured $H_0$ is the correct value, but only treat 
it as an adjustible parameter in a standard multiparameter fit that 
does not include the local measurement of $H_0$, 
the standard $\L$CDM model is doing better \cite{cline} 
than our modified model. Our claim that to some extent our modified 
Friedmann equation ease the $H_0$ tension, is the statement  that 
it can incorporate the locally measured value of $H_0$ as a data point, 
and still fit the other 
late time cosmology observations with reasonable accuracy. } (see 
\cite{fit1,fit2,mod3}. This is the value obtained
by local measurements \cite{local} 
and the value that gives rise to the so-called $H_0$ tension, 
since the best value for $H_0$ obtain by fitting (non-local data) 
to the $\L$CDM model is 67 \versionII{km/s/Mpc} \cite{planck}. 
The difference is around 5-6 $\sigma$.  
The late time effective Hamiltonian $H_{\rm lt}$ we consider is thus\footnote{
\versionII{
In eq.\ (62) we have added CDM to our Hamiltonian $H_0$ by hand 
and in an entirely classical way. 
One could indeed ask if matter should not have been included 
as quantum matter at an earlier stage. 
Presently we do not know how to do this.
}
} 
\beq\label{g3} 
 H_{\rm lt}(V,P) =
 V \Big(\! - \frac{3}{4} \big(P^2 +  \frac{2g}{P} \big) + 
 \k_3 \rho_{\rm m}(V) \Big),
\eeq
where $\rho_{\rm m}(V)$, representing the CDM energy density, 
will satisfy
\beq\label{g4}
  \k_3 \rho_{\rm m}(V(t))
 = \frac{C}{V(t)}, \quad C= V(t_0) \k_3 \rho_{\rm m}(t_0),
\eeq
where $V(t_0)$ and  $\rho_{\rm m}(t_0)$ denote 
the ``volume'' (eq.\ \rf{g2}) and 
the CDM density at the present time $t_0$. 
By energy conservation $V \rho_{\rm m}(V)$ will not appear 
in the eom related to $H_{\rm lt}(V,P)$,
only in expressions that involve the energy, 
like the (modified) Friedmann  equation, 
which is simply $H_{\rm lt}(V,P) = 0$. 
The eoms can now be written as the Friedmann equation 
and an equation for the time dependence of 
the Hubble parameter $H(t) = \dot{a}/a = 1/3\; \dot{V}/{V}$, 
and a (not indepent) equation for $\dot{P}$, 
where $\dot{x}$ denotes the derivative with respect to time $t$ 
associated with $H_{\rm lt}(V,P)$:
\beq\label{g5}
  \k_3 \rho_{\rm m}{\negtinyspace}\big(V(t)\big)
  = \frac{3}{4}\Big( P^2 + \frac{2g}{P} \Big), 
  \quad 
  H(t) = \frac{\dot{V}}{3V} = 
  \oh \Big(\! -P + \frac{g}{P^2} \Big), 
  \quad 
  \dot{P} = \k_3 \rho_{\rm m}{\negtinyspace}\big(V(t)\big)
  .
\eeq
The last equation is derived from the preceding two equations 
together with 
\rf{g4} and can readily be integrated, 
allowing us to express $t$ as an elementary function of $P$ 
(see \cite{mod3} or eq.\ \rf{g11} below). 
It is convenient to rewrite \rf{g5} in dimensionless variables. We introduce 
\beq\label{g6}
  \l = \frac{4}{9} \frac{ \k_3 \rho_{\rm m}}{ (\sqrt{2} g)^{2/3}}, \quad h = 
  \frac{2\sqrt{2}}{3}\frac{H}{ (\sqrt{2}g)^{1/3}},
  \quad p = \frac{P}{(\sqrt{2} g)^{1/3}},
  \quad \tau = \frac{9}{4} (\sqrt{2} g)^{1/3} {\tinyspace} t.
\eeq
Eqs.\ \rf{g5} can then be written as 
\beq\label{g7}
  \l(p) = \frac{1}{3} \Big(p^2 + \frac{\sqrt{2}}{p}\Big),
  \quad
  h(p) = \frac{1}{3} \Big(\!\! -\!\sqrt{2} p + \frac{1}{p^2} \Big),
  \quad
  \frac{\dd p}{\dd \tau} 
  = \l(p),
\eeq
and it is seen that 
\beq\label{g8}
  \l(p) = h(-1/p)\quad {\rm and} \quad
  f(\l,h) := 2\l^3 -3 \l^2h^2 + 2 h^3 -1 = 0.
\eeq 
We can thus view the solution $\l(p(\tau)),h(p(\tau))$ 
to the modified Friedmann equation as lying on the algebraic curve 
defined by the second equation in \rf{g8}. 
This algebraic curve (over the real numbers) consists of two branches,
corresponding to $p\in ]-\infty,0[$ and $p\in ]0,\infty[$, 
as shown in Fig.\ \ref{fig4}.

The two branches intersect in a singular point 
or node in the framework of algebraic curves, and 
this point has to be $(\l_s,h_s) =(1,1)$ 
since $ \l(p) = h(-1/p)$. 
In this singular point we have 
\beq\label{g9}
 f(\l,h) =0,\quad  \frac{\prt f}{\prt h} = \frac{\prt f}{\prt \l} =0,
 \quad {\rm for} \quad
 (\l,h) = (\l_s,h_s).
\eeq
The solution $\l(p(\tau)),h(p(\tau))$ representing an expanding universe, 
starting from $\tau =0$ at $v(\tau=0)=0$, i.e.\ 
$\l(\tau=0) = \infty$, and expanding exponentially at late time, 
where $\l(p) \to 0$ and $p \to -2^{1/6}$,
is located on the part of the negative $p$ branch of the algebraic curve 
where $p \in ]-\infty, -2^{1/6}[$.  
It includes the value of the singular point,
$p_s = -(1+ \sqrt{3})/\sqrt{2}$.  
As already mentioned the relation between $\tau$ and $p$ 
can be found in terms of elementary functions 
by solving the last equation in \rf{g7}. 
It can be written as 
\beq\label{g10}
  2^{1/6} \dd \tau = \frac{3 {\tinyspace} \dd z}{1-z^3} = 
  \sum_{k=0}^2 \frac{\dd z}{1- \om_k z},
  \quad
  z= - \frac{2^{1/6}}{p}, \quad 
  \om_k=\e^{\frac{2\pi i}{3} k},
  \eeq
  \beq\label{g11}
 2^{1/6} \tau = \int_0^{z^3}\!\! \frac{\dd y}{y^{2/3} (1-y)} = B(z^3;1/3,0)=
 -\sum_{k=0}^2 \frac{1}{\om_k} \ln(1- \om_k z),
\eeq 
where $B(x;a,b)$ denotes the incomplete beta-function and $\om_k$ are 
the three solutions to $z^3=1$. $\tau \to 0$ corresponds to $z \to 0$, 
i.e.\ $p \to -\infty$,
while $\tau \to \infty$ corresponds to $z \to 1$, i.e.\ $p \to -2^{1/6}$. 
At the singular point the value 
$\tau_s = 2^{-1/6} B(z_s^3; 1/3,0)$, $z_s = -2^{1/6}/p_s$,
i.e.\  $\tau_s = 1.639$.
 
The fact that   $\l_s= h_s =1$ and $\tau_s = 1.639$ can be translated 
back to ``physical'' units via \rf{g6} and we obtain
\beq\label{g11a}
 \sqrt{\k_3 \rho_{{\rm m},s}} = \frac{3}{2} (\sqrt{2} g)^{1/3},\quad 
 H_s = \frac{3}{2 \sqrt{2}} (\sqrt{2} g)^{1/3},
 \quad t_s = \frac{4}{9}{\tinyspace} \tau_s (\sqrt{2} g)^{-1/3}.
 \eeq
 Thus we can write, without any reference to the value of $g$:
 \beq\label{g11b}
 \Omega_{{\rm m},s} = \frac{\k_3 \rho_{{\rm m},s}/3}{H_s^2} = \frac{2}{3},
 \qquad 
 \frac{\Length_s}{t_s} = \frac{1/H_s}{t_s} = \frac{3}{\sqrt{2}\,\tau_s}
 = 1.294,
\eeq
where at the singular time $t_s$,
the former $\Omega_{{\rm m},s}$ denotes the matter density parameter,
whereas the latter $\Length_s/t_s$ represents
the characteristic expansion speed of the Universe.
We will comment on these relations in the discussion section. 
  
Above we considered $f(\l,h)=0$ as an algebraic curve over the real numbers.
We can generalized it to an algebraic curve over the complex numbers 
(i.e.\ a surface). 
In this case there are three singular points, namely 
\beq\label{g12}
 (\l_s,h_s) \ = \ (\om_0,\om_0), \quad (\om_1,\om_2), \quad (\om_2,\om_1),
\eeq
and the complex algebraic curve becomes amazingly simple if we view it
as a section of a projective algebraic curve. We introduce projective 
coordinates $(x,y,z)$ by writing 
\beq\label{g13}
 \l = \frac{x}{z},
 \quad
 h= \frac{y}{z},
 \qquad
 F(x,y,z)
 = z^4 f\Big(\frac{x}{z},\frac{y}{z}\Big)
 =
 2x^3z-3x^2y^2 + 2y^3z -z^4,
\eeq
and the projective modified Friedmann equation is now $F(x,y,z)=0$. 
It can be simplified by performing a basis change:
\beq\label{g14}
 x = \om_1 X + \om_2 Y +  Z, \quad y = \om_2 X+ \om_1 Y +Z,
 \quad z = X+Y+Z,
\eeq
such that the projective modified Friedmann equation reads:
\beq\label{g15}
  -{\tinyspace}\frac{1}{27} F(x,y,z) = X^2 Y^2 + Y^2 Z^2 + Z^2 X^2 = 0,
\eeq
and the singular points are
\beq\label{g16}
  (X,Y,Z) \ = \ (0,0,1), \quad (1,0,0), \quad (0,1,0).
\eeq
The point (0,0,1) is the one corresponding to 
the physical singular point $(\l_s,h_s) =(1,1)$, 
the only one of the singular points that corresponds to 
real values of $\l$ and $h$. 
Returning to the algebraic curve over the real numbers,
the only point that have a special status is $(\l_s,h_s) =(1,1)$. 
We will discuss the implication of this below. 
Here we can ask if there is a symmetry 
that fixes the value of $(\l_s,h_s)$ to be $(1,1)$. 
As we noted above it was fixed to the specific value 
due to the first equation in \rf{g8}: $\l(p) = h(-1/p)$. 
Considering $p$ as a complex value 
we can relate it to an $\mathrm{SL}(2,\Dbl{C})$ transformation as follows. 
First we introduce the Virasoro generators by 
\beq\label{g17}
 \ell_n := - p^{n+1} v,
 \quad
 \{ \ell_n , \ell_m \} = (n-m) \ell_{n+m}, 
\eeq
where $\{\cdot,\cdot\}$ denotes 
the Poisson bracket for the classical $(v,p)$ system. 
We can write $\ell_n$ as an operator $\hat{\ell}_n$ 
acting on functions $f(p)$ in the complex plane as follows
\beq\label{g18}
 \hat{\ell}_n f(p) := \{ \ell_n , f(p) \}
 = -p^{n+1} \frac{\dd f}{\dd p}, 
 \quad
 [\, \hat{\ell}_n , \hat{\ell}_m \,]  = (n-m) \hat{\ell}_{n+m}.
\eeq
$\hat{\ell}_{-1},\hat{\ell}_0,\hat{\ell}_1$ are 
the standard generators of $\mathrm{SL}(2,\Dbl{C})$ 
transformations and the exponential of these are 
the globally defined conformal transformations in the complex $p$ plane:
\beq\label{g19}
 \e^{a \hat{\ell}_{-1} } f(p) = f(p-a),
 \quad
 \e^{a \hat{\ell}_{0} } f(p)  = f(\e^{-a} p),
 \quad 
 \e^{a \hat{\ell}_{1} } f(p)  = f\big(p/(1+ap)\big).
\eeq
In particular we have 
\beq\label{g20}
\hat{g} 
= e^{ \hat{\ell}_{-1} }e^{ \hat{\ell}_{1} }e^{ \hat{\ell}_{-1} },
\quad
\hat{g} f(p) 
= f(-1/p),\quad \hat{g} \in \mathrm{SL}(2,\Dbl{C}).
\eeq
Thus
\beq\label{g21}
\hat{g} \l (p) = \l (-1/p) = h(p), \quad  \hat{g} h(p) = h (-1/p) = \l(p),
\eeq
and the relation \rf{g8} can be understood as the algebraic curve 
related to the modified Friedmann equation allowing a $\mathrm{SL}(2,\Dbl{C})$ 
transformation $\hat{g}$ such that $\hat{g} \l = h$, $\hat{g} h = \l$.

\section{\versionII{Cosmological relations emerging from the model}}\label{predictions}

Below we discuss possible implications of the model described above.
\versionII{
Although part of the analysis is necessarily approximate 
and the calculations done below will use a number of approximations 
and assumptions, we nevertheless are of the opinion that 
the estimates done are going beyond a purely qualitative analysis.
}



\subsection{Large numbers in our model}
\label{sec:LargeNumbers}

 We have seen that the modified Friedmann equation defines 
a special singular point when we view the solution 
as part of an algebraic curve. 
Expressed in dimensionless units, 
like the ones introduced in \rf{g6} this special point is 
\beq\label{k1}
 (\l_s,h_s) = (1,1),
\eeq
and it is a natural reference point for the modified Friedmann equation. 
We will now argue that the coincidence of various ratios of 
physical observables related to our Universe can be understood 
if present Universe is in the 
(not necessarily very close) neighborhood of this singular point.

The ratio between the current time $t_0$ and 
the Planck time $t_{{\rm planck}}$ 
is approximately
\begin{equation}\label{CurrentTimeVsPlanckTime}
\frac{t_0}
     {t_{{\rm planck}}}
\,\sim\,
10^{60}
\,.
\end{equation}
Similarly, the ratio between the  current size of universe, i.e.\ 
the current Hubble radius $\Length_0= 1/H_0$, 
and the Planck length 
$\Length_{{\tinyspace}{\rm planck}}\;\;(= \sqrt{G_{\rm N}}$) is
\begin{equation}\label{CurrentUniverseSizeVsPlanckSize}
\frac{\Length_0}
     {\Length_{{\tinyspace}{\rm planck}}}
\,\sim\,
10^{60}
\,,
\end{equation}
i.e.\ of the same order of magnitude.
In addition 
we have that the ratio of the total matter energy of Universe 
$E_{{\halftinyspace}{\rm univ}}$ 
to the Planck energy 
$E_{{\halftinyspace}{\rm planck}}= 1/\sqrt{G_{\rm N}}$ is 
\begin{equation}\label{UniverseEnergyVsPlanckEnergy}
\frac{E_{{\halftinyspace}{\rm univ}}}
     {E_{{\halftinyspace}{\rm planck}}}
\,\sim\,
10^{60}
\,.
\end{equation}
The value of the current matter energy density 
$\rho_{{\rm m},0}$ 
to the Planck matter energy density 
$\rho_{{\halftinyspace}{\rm planck}}$
(defined as 
$\k_3 \rho_{{\halftinyspace}{\rm planck}} = 
 \k_3 E_{{\halftinyspace}{\rm planck}}/(\Length_{{\tinyspace}{\rm planck}})^3
 = 
1/G_{\rm N}^2 $) is
\begin{equation}\label{CurrentUniverseEnergyDensityVsPlanckEnergyDensity}
\frac{\rho_{{\rm m},0}}
     {\rho_{{\halftinyspace}{\rm planck}}}
\,\sim\,
10^{-120}
\,.
\end{equation}
These four relations are not independent. 
As an example one can derive \rf{UniverseEnergyVsPlanckEnergy} 
from 
\rf{CurrentUniverseEnergyDensityVsPlanckEnergyDensity}
and 
\rf{CurrentUniverseSizeVsPlanckSize} as 
\begin{equation}
E_{{\halftinyspace}{\rm univ}}
\,\sim\,
\rho_{{\rm m},0}{\dbltinyspace}
 \Length_0^3
\qquad\hbox{and}\qquad
E_{{\halftinyspace}{\rm planck}}
\,\sim\,
\rho_{{\halftinyspace}{\rm planck}}{\dbltinyspace}
 \Length_{{\tinyspace}{\rm planck}}^3
\,.
\end{equation}

The current time $t_0$ 
and the current value of the Hubble parameter $H_0$
may at first glance appear to be values unrelated in the theory. 
However, their values at the singular point of the algebraic curve 
{\it are related} since we in dimensionless variables have 
$\l_s = 1 = h_s$ for $\tau_s = 1.639$. 
As we saw in eqs.\ \rf{g11b} 
the ratio between $t_s$ and $\Length_s$ is of order 1,
precisely as ratio between the experimental values $t_0$ and $\Length_0$, 
according to 
\rf{CurrentTimeVsPlanckTime} and \rf{CurrentUniverseSizeVsPlanckSize}.
Similarly the ratio between 
$1/\sqrt{\k_3 \rho_{{\rm m},s}}$ and $\Length_s$ is 
of order 1 according to 
\rf{g11b}, and the same is the case for the observed ratio between 
$1/\sqrt{\k_3 \rho_{{\rm m},0}}$ and $\Length_0$. 
Provided that our late time cosmology 
is described by the modified Friedmann equation it then tells us that 
the present time $t_0$ is not too far from the time $t_s$ related to the 
singular point of the algebraic curve for the modified Friedmann equation.

\subsection{The smallness of the coupling constant $g$}
\label{sec:smallg}

The ratios in eq.\ \rf{g11b} were independent of 
the actual value of the coupling constant $g$. 
However, looking at the relations \rf{g11a} it is clear that 
the actual values of $t_s$, $\Length_s$ and $1/\sqrt{\k_3 \rho_{{\rm m},s}}$ 
will be of order $O(g^{-1/3})$. 
According to the arguments just given, the same should then 
be true for $t_0$, $\Length_0$ and $1/\sqrt{\k_3 \rho_{{\rm m},0}}$, 
but these values are $10^{60}$ 
(times the Planck length $\Length_{\rm planck} = \sqrt{G_{\rm N}}$\,) 
according to 
\rf{CurrentTimeVsPlanckTime},  \rf{CurrentUniverseSizeVsPlanckSize} and 
\rf{CurrentUniverseEnergyDensityVsPlanckEnergyDensity}.
We thus obtain that 
\beq\label{xx0}
g^{-1/3} \sim 10^{60}   \sqrt{G_{\rm N}}.
\eeq

In \cite{mod1,mod2,mod3} we determined 
from cosmological  observations the value of $g$ 
that gave the best fit to the data, 
and we found indeed that  $g^{-1/3} \sim 10^{60} \sqrt{G_{\rm N}}$.

To conclude: 
if our late time cosmology is described by the modified Friedmann equation, 
being presently in the neighborhood of the singular point 
associated with the associated algebraic curve 
allows us to ``understand'' some of the hierarchy problems in cosmology, 
namely
\begin{itemize}
\item
the time hierarchy 
(the ratio of the present time $t_0$
 to the Planck time $t_{\rm planck}$), 
\item
the size hierarchy 
(the ratio of the size $\Length_0$ of the present Universe
 to the Planck size $\Length_{\rm planck}$),

\item
the energy density hierarchy of matter 
(the ratio of the present-day energy density of matter $\rho_{{\rm m},0}$
 to the Planck energy density $\rho_{\rm planck}$).
\end{itemize}
The value of $g$ needed to fit observations seems to be very small 
($g^{1/3}\sqrt{G_{\rm N}} \sim 10^{-60}$\,), 
in the same way as the value of the cosmological constant $\L$ 
in the $\L$CDM model is also widely considered unnaturally small 
($\sqrt{\L}\sqrt{G_{\rm N}} \sim 10^{-60}$\,) . 

In our case it is natural to ask 
if the knitting and exchange mechanisms needed 
to get from the underlying $W_3$ algebra to 
the effective higher dimensional modified Friedmann equation might provide 
an explanation for the smallness of the $g$ encountered in this equation. 
\versionII{
Admittedly, the calculations presented below assume that 
tree-diagrams dominate the knitting process, 
that the CDT propagators mediating the knitting can be 
approximated by their short distance approximation 
and that in this short distance approximation one further approximates 
 the time-extent of wormhole propagator, 
$t_{\rm wh}$, to be  $1/d$ times the total time it takes to create 
a $d$-dimensional knitting-web. 
While one can argue for each of these approximations, 
they are of course at the level of assumptions. With these reservations made, let us proceed.}

Loosely speaking, points in the higher-dimensional space are formed 
by knitting together points from lower one-dimensional spaces. 
One would naively expect them to be further apart than 
in the original one-dimensional spaces.
We thus write 
\beq\label{s1}
  L^{(k)} = \g L_0^{(k)} ,
  \qquad\hbox{[\,$k\!=\!1$, $2$, \ldots, $d$\,]},
\eeq
where 
$L^{(k)}$ [\,$k\!=\!1$, $2$, \ldots, $d$\,] 
denotes length in the higher dimensional space 
and 
$L_0^{(k)}$ [\,$k\!=\!1$, $2$, \ldots, $d$\,] 
length as it appears in the original one-dimensional spaces. 
We here define the $d$-dimensional volumes $V$ and $V_0$ by 
\begin{equation}
V \define \frac{1}{\k_d} L^{(1)} L^{(2)} \ldots L^{(d)}
\,,\qquad
V_0 \define \frac{1}{\k_d} L_0^{(1)} L_0^{(2)} \ldots L_0^{(d)}
\,.
\end{equation}
Let us write the amplitude 
for the process depicted in Fig.\ \ref{fig2}. 
We denote the fields before knitting the bare fields.
\beq\label{s10}
  \Phi^{\rm (bare)} (\{V_0\}) =
  \Psi^{\rm (bare)}(L_0^{(1)}) \cdots \Psi^{\rm (bare)}(L_0^{(d)}),
  \quad
  \{V_0\} \define \{L_0^{(1)},\ldots,L_0^{(d)}\} ,
\eeq
 and consider the amplitude
\beq\label{s11}
  A = \vac \Phi^{\rm (bare)}(\{V_0'\}) \, 
  \e^{- \T (\hH_0^{\rm (bare)} + \hH_{\rm wh}^{\rm (bare)})}
  \Phi^{\dg {\rm (bare)}}(\{V_0\}) \cuum.
\eeq
Here $\hH_0^{\rm (bare)}$ is the quadratic part of Hamiltonian 
coming from the $W_3$ algebra, i.e.\ for each flavor
it is given by terms  like in eq.\ \rf{g1}, 
dressed by the second quantized fields $\Psi^{\rm (bare)}(L_0^{(k)})$. 
The conjecture is now that 
the wormhole interaction governed by $\hH_{\rm wh}^{\rm (bare)}$ 
will effectively leave us with a $d$-dimensional space, 
where we have a $d$-dimensional Hamiltonian, 
like the one referred to in \rf{g2} in the three-dimensional case. 
%
%
However, we assume it will have the same functional form 
in any spatial dimension $d$ resulting from knitting. 
We appeal here to maximal symmetry and the fact that 
the Hartle-Hawking minisuperspace Hamiltonian as written in \rf{g2} 
has the same functional form in any $d \!>\! 1$. This assumption implies that 
also the effective  constant $G_{\rm N}$ of dimension $d\!-\!1$ 
that appears in \rf{g2} (in the case of $d\!=\!3$) 
is created during knitting. Two things appear simultaneously: an effective
$d$-dimensional Hamiltonian  of the form \rf{g2} is created 
and the length scale is changed by a factor $\g$. 
If we artificially separate this creation  in 2 steps then we first obtain
\beq\label{xx1}
 H(V_0,P_0,g_0) \propto  -\, V_0 \Big( P_0^2 - \mu_0+ \frac{2g_0}{P_0}\Big),
\eeq
when we change the one-dimensional Hamiltonian \rf{g1} to 
a $d$-dimensional Hamiltonian.
Then by the change of scale factor \rf{s1} we obtain
\beq\label{xx2}
 H(V_0,P_0,g_0)  \to \g^d H(V,P,g), 
\eeq
where 
\beq\label{xx3}
  V = \g^d V_0, \quad P= \g^{-d} P_0, 
 \quad \mu = \g^{-2d} \mu_0, \quad g = \g^{-3d} g_0.
\eeq
In the new $d$--dimensional variables we write 
\bea\label{xx4}
A
&=&
\la V_0' | \, e^{- t H(V_0,P_0,g_0)} | V_0\ra
=
\la V' | \, e^{-\gamma^d t H(V,P,g)} | V\ra
\nonumber\\
&=:&
\frac{\k_d V_0}{L_{\rm unit}^d} A_\ast
\,,
\qquad
A_\ast \approx \gamma^d
\,,
\eea
where we assumed that 
$\T = \T_{\rm unit} \sim L_{\rm unit}$. 
The factor $\gamma^d$ arises 
because a single Planck--scale wormhole web occupies 
a $d$-dimensional Planck cell of volume $L_{\rm unit}^d$, 
so that the rescaling of lengths by $\gamma$ enhances 
the local quantum weight by $\gamma^d$.
Here $A_\ast$ denotes the quantum amplitude 
associated with a single Planck--scale wormhole web.
A wormhole web is a dense network of 
parametrically short cylindrical wormholes, 
which fills the emergent $d$--dimensional space 
and forms a spacetime condensate.
Each spacetime cell of volume $L_{\rm unit}^d$ is occupied by one such web,
so that the total amplitude is proportional to the number of cells,
$\k_d V_0/L_{\rm unit}^d$.

A wormhole web, 
shown as the magenta tree graph 
in Fig.\ \ref{fig2}, 
consists of parametrically short cylindrical wormholes 
connecting one-dimensional universes. 
Each thick magenta segment corresponds to a one-dimensional universe, 
and each node represents a Planck-scale spatial cell. 
At the quantum level, the web is characterized 
by the local amplitude $A_\ast$ (Eq.\ \rf{xx4}), 
which gives the weight of a single Planck--scale web 
inside the knitted spacetime.
Physically, this means that at each spacetime point
there exists a Planck--scale wormhole web.
When the knitted spacetime is viewed as 
an emergent $d$--dimensional continuum,
these local quantum weights are accumulated over all Planck--size cells.
We introduce the temporal extent of a wormhole web, $t_{\rm unit}$,
which we immediately identify with the Planck time $t_{\rm planck}$,
and correspondingly $L_{\rm unit}=L_{\rm planck}$.

This structure should be contrasted with ordinary two--dimensional CDT,
where the cosmological constant weights geometries by a factor
$e^{-\lambda V}$.
In the knitted $(d\!+\!1)$--dimensional spacetime, the corresponding
cosmological term instead counts the density of wormhole webs,
so that it weights amplitudes by the number of Planck--scale cells
threaded by the condensate.
It is therefore not an ordinary vacuum energy,
but a measure of how strongly spacetime is populated by wormhole webs.
This expansion to lowest order in 
$\T = \T_{\rm unit} \sim L_{\rm unit}$ 
is appropriate after the knitting process, 
where one must work in units of $t$---the timescale of 
the knitting process itself. 
This timescale will be related to 
the dynamically generated gravitational constant $G_{\rm N}$. 
Consequently, the matrix element $| \la V'| t H(V,P,g) | V\ra$ 
can be considered of order unity, 
implying that $A$ is a pure dimensionless number.
Eq.\ \rf{xx4} allows us to write eq.\ \rf{xx3} as 
\beq\label{xx5}
  \mu = A_\ast^{-2} \mu_0 ,
  \qquad
  g = A_\ast^{-3} g_0 ,
  \qquad
  L^{(k)} = A_\ast^{1/d} L_0^{(k)} ,
  \quad\hbox{[\,$k\!=\!1$, $2$, \ldots, $d$\,]}.
\eeq

It now remains to estimate the factor $A_\ast$. 
An estimate comes from the calculation of the wormhole propagators 
shown in Fig.\ \ref{fig2}, 
these being responsible for the knitting of 
the $d$ one-dimensional spatial universes to a $d$-dimensional universe.
These are the original CDT propagators 
$\Delta(L_{\rm wh},L'_{\rm wh}; \T_{\rm wh})$ \cite{al1}, 
where subscript ``wh'' refers to ``wormhole''. 
As already mentioned these propagators are quite complicated, 
but in a certain limit where $\T_{\rm wh}$ is small, 
they can be approximated by 
$\Delta \sim \sqrt{L_{\rm wh}/\T_{\rm wh}}$\,\footnote{
This is the cylinder amplitude of very short height $\T_{\rm wh}$, 
with each of its two boundaries of length $L_{\rm wh}$ 
and a marked point on each.}. 
Consider a $d$-dimensional space, created as shown in Fig.\ \ref{fig2}, 
from $d$ one-dimensional spaces, connected by tree-graphs. 
Such a tree-graph will have 
$N_{\rm vertex} = d-2$ internal and 
$N_{\rm edge} = d$ external vertices, respectively, 
and thus $N_{\rm wh} = 2d-3$ internal lines. 
The internal lines represent wormhole propagators, 
which in our approximation will contribute with 
$\Delta^{\!N_{\rm wh}} \sim  (L_{\rm wh}/\T_{\rm wh})^{(2d-3)/2}$. 
We have to multiply this contribution with the number of tree-diagrams 
with $d$ external vertices and in addition with 
the number of permutations of the $d$ flavors 
of the external vertices of the tree-diagrams. 

Before estimating the factor $A_\ast$, we must specify the physical scale 
that sets the typical wormhole length.
The Planck length $L_{\rm planck}$ 
and the corresponding Planck energy $L_{\rm planck}^{-1}$ 
should be regarded as derived quantities, 
generated by the knitting process from the bare parameters 
$\mu_{\rm wh} \!\define\! \mu_0$ and $g_0$. 
According to \rf{e15}, 
the characteristic wormhole length scale is set by 
the wormhole cosmological constant,
\begin{equation}\label{WormholeSize}
L_{\rm wh} \,\sim\, \frac{1}{\sqrt{\mu_{\rm wh}}} ,
\end{equation}
and we identify this dynamically generated scale with the Planck length,
\begin{equation}\label{WormholeWebSize}
L_{\rm planck} \,\define\, L_{\rm wh} .
\end{equation}
All subsequent estimates of $A_\ast$ will be expressed 
in these dynamically generated units.
Totally we then obtain a contribution
\beq\label{s2}
  A_\ast \approx
  (2d-5)!!
  \left(\frac{L_{\rm wh}}{\T_{\rm wh}}\right)^{\!\!(2d-3)/2}
   ( g_0 L_{\rm wh}^3 )^{2(d-1)}
  \,,
\eeq
where 
$2d\!-\!3 = N_{\rm wh}$ is the number of internal lines, 
$2(d\!-\!1) = N_{\rm vertex} \!+\! N_{\rm edge}$ is 
the number of vertices and edges, 
$(2d\!-\!5)!!$ is the number of tree diagrams. 
The approximation 
$\Delta(L_{\rm wh},L'_{\rm wh}; \T_{\rm wh}) \sim 
 \sqrt{L_{\rm wh}/\T_{\rm wh}}$
assumes that $L_{\rm wh},L'_{\rm wh}$ is $O(1)$ and that 
$\T_{\rm wh} \ll L_{\rm wh}$, but not too small. 
A natural choice is 
$\T_{\rm wh} \sim  \T_{\rm unit}/d$
and 
$\T_{\rm unit} = \T_{\rm planck} \sim L_{\rm wh}$, since 
the total time related to the tree-graph is of order $L_{\rm wh}$,
and the endpoints of the tree-graph, 
each of which connects to a flavor-carrying space, 
partition the total time $\T_{\rm unit}$ into $d$ segments.
%
Inserting this in \rf{s2} we obtain for $d\!=\!25$
\beq\label{s3}
 A_\ast \approx
 \versionII{2 \times}
 10^{61}
 ( g_0 L_{\rm planck}^3 )^{48}
 \,.
\eeq
Here we have used 
$L_{\rm wh}=L_{\rm planck}$. 
\versionII{
When comparing to experiments we assume $g_0 L_{\rm planck}^3 \approx 1$. 
}

In a similar vein, 
starting from a bare cosmological constant 
$\mu_0 L_{\rm planck}^2 = O(1)$, 
the knitting mechanism produces a vacuum energy of Planckian order, 
which is then converted into the big-bang energy. 
The effective gravitational coupling, however, 
is suppressed by the same large factor, 
leading to the hierarchy \rf{UniverseEnergyVsPlanckEnergy}.

In our effective cosmological description, 
the ordinary cosmological constant is assumed to vanish 
by the Coleman mechanism, 
so the small observed value is 
not attributed to the wormhole-web sector 
but to effects beyond it.  

Also, it was important that we had $d\!=\!25$ and not, e.g., $d\!=\!3$. 
The choice $d\!=\!25$ corresponds to the number of flavors 
in the $H_3(\mathbb{O})$ model, 
which naturally leads to the desired hierarchy factor. 
This model thus appears as the most interesting one, 
and, as described above, 
by symmetry breaking combined with the exchange mechanism, 
it can produce models where the dimension of the extended space 
ranges from $9$ down to $3$. 

It should be noticed that the ``large'' value $d\!=\!25$ also implies 
that the scale factor $\g$ appearing in Eq.\ \rf{s1} 
is not outrageously large. 
Using the estimate of $A_\ast$ given in Eq.\ \rf{s3} 
together with the approximate relation $A_\ast \sim \gamma^d$ 
from Eq.\ \rf{xx4}, and taking $g_0 L_{\rm planck}^3 \approx 1$, 
we then find 
\beq\label{s4}
\g \approx  \versionII{280}
\,.
\eeq

Finally, let us examine the connection 
between the Planck--scale wormhole--web amplitude $A_\ast$ 
and the large--scale energy budget of the Universe.
Here and in the following, 
$D$ denotes the number of macroscopic spatial dimensions of the Universe 
selected by the Coleman mechanism, which in general satisfies 
$D \!\le\! d$. 
(In the case of our Universe, $D \!=\! 3$.)
Through the knitting mechanism, 
the microscopic quantum weight $A_\ast$ of 
a single Planck--scale wormhole web is promoted to an effective
cosmological term in the emergent higher--dimensional spacetime.
Because a web occupies every Planck--scale cell,
the total two--dimensional amplitude grows proportionally
to the number of such cells, i.e.\ to the spatial volume,
and therefore acts as a cosmological constant in the knitted theory.

\begin{figure}
\centerline{ \includegraphics[width=400pt]{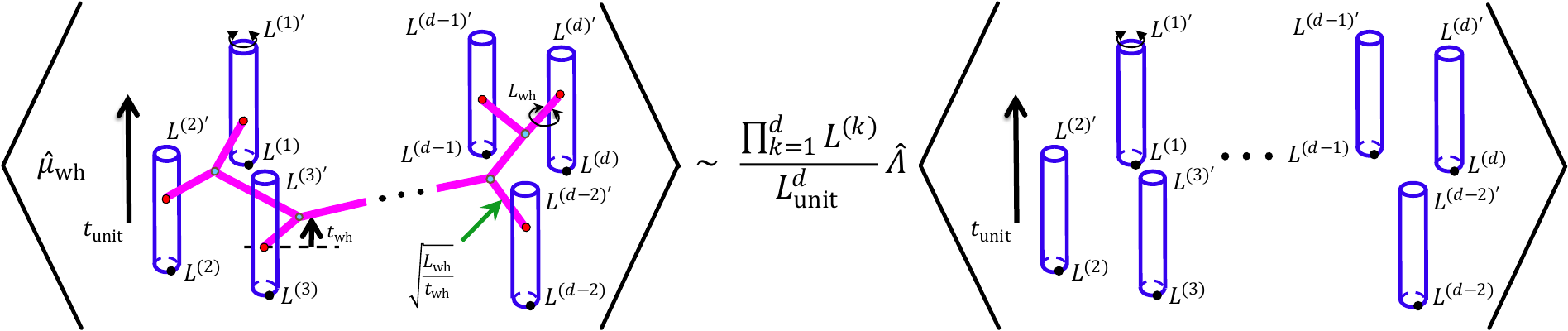}}
\caption{{\footnotesize
The graphical representation of Eq.~\rf{KnittingCC0},
which describes the knitting mechanism
from the viewpoint of the wormhole web
(magenta tree graph).
Here each thick magenta segment represents a cylindrical
one--dimensional universe connected through wormholes,
even though it is drawn as a line in the figure.
The full magenta network corresponds to a single 
wormhole web with quantum weight $A$.
The Planck--scale wormhole web occupies a spacetime cell of 
temporal extent $\T_{\rm unit}$ and spatial size $L_{\rm unit}$,
both of which are dynamically generated by the
two--dimensional quantum gravity dynamics.
In particular,
$\T_{\rm unit} = \T_{\rm planck} \sim L_{\rm wh}$ 
and
$L_{\rm unit} = L_{\rm planck} \sim L_{\rm wh} = 1/\sqrt{\mu_{\rm wh}}$,
so that the web lives in a Planck--scale spacetime cell.
%
$\mu_{\rm wh}$ denotes the wormhole cosmological constant.
The numbers of vertices, edges, and internal lines in the magenta
tree graph are
$N_{\rm vertex}=d-2$,
$N_{\rm edge}=d$,
and $N_{\rm wh}=2d-3$,
respectively.
}}
\label{figWormholeWebEnergy}
\end{figure}

We assume that tree graphs built from 
parametrically short cylindrical segments provide 
the dominant contribution among all wormhole--web diagrams.
Figure~\ref{figWormholeWebEnergy} illustrates this matching:
the magenta tree graph represents a single wormhole web 
with quantum weight $A$,
while the factor $\k_d V_0/L_{\rm unit}^d$ counts how many such
cells fill the knitted spacetime.
This correspondence is encoded in
\begin{equation}\label{KnittingCC0}
\versionII{\hat\mu_{\rm wh}} A
\,\sim\,
\frac{\k_d V_0}{L_{\rm unit}^d}\,\versionII{\hat\Lambda}
\,,
\end{equation}
and therefore for a single Planck--scale web
\begin{equation}\label{KnittingCC}
\mu_{\rm wh} A_\ast
\,\sim\,
\Lambda
\,,
\end{equation}
where $\mu_{\rm wh}$ and $\Lambda$ denote the cosmological constants
associated with the wormhole and the emergent $d$--dimensional space,
respectively.
The corresponding vacuum energies associated with a fundamental cell
of characteristic size $L_{\rm unit}$ are defined by
$\hat\mu_{\rm wh}:=L_{\rm unit}\mu_{\rm wh}$
and
$\hat\Lambda:=(L_{\rm unit}^d/\k_d)\Lambda=L_{\rm unit}\Lambda$.
Since $\mu_{\rm wh}>0$, Eq.~\rf{KnittingCC}
implies $\Lambda>0$.
This suggests that the emergent geometry is of de Sitter type.

Equation~\rf{KnittingCC0} expresses the correspondence between
the vacuum energy carried by the wormhole web and the vacuum energy
assigned to the Planck--scale cells of the emergent spacetime.
Equation~\rf{KnittingCC} is the corresponding relation for a single
Planck--scale web and can be written as
$\Lambda \sim A_\ast \mu_{\rm wh}$. 
This relation shows that the cosmological constant in the emergent
$d$--dimensional spacetime is not a fundamental vacuum energy,
but a statistical quantity determined by the wormhole web.
Each Planck--scale cell of the knitted spacetime can be realized by
a large number of microscopic wormhole--web configurations,
and $A_\ast$ counts their combinatorial weight.
The effective cosmological constant $\Lambda$ therefore measures
the statistical abundance of Planck--scale wormhole webs
in the emergent spacetime.
In this sense, the cosmological constant is the macroscopic
manifestation of the microscopic knitting process.
%

The quantity $\Lambda$ plays the role of a vacuum--energy density 
in the effective $d$--dimensional gravitational equations, 
but physically it represents the 
\versionII{statistical} density of Planck--scale wormhole webs 
in the knitted spacetime.
Although the wormhole web is constructed by knitting together $d$
one--dimensional universes, this does \textit{not} imply that 
the emergent spacetime has $d$ macroscopic spatial dimensions.
In general, only $D \!\le\! d$ directions become extended, 
while the remaining $d\!-\!D$ directions are dynamically stabilized 
at the Planck scale.
Consequently, the total knitted volume factorizes as
\begin{equation}
\k_d V_0 \;\sim\; \k_D V_D\,L_{\rm planck}^{\,d-D},
\end{equation}
where $V_D$ is the physical $D$--dimensional spatial volume.
Since each wormhole web occupies a single $d$--dimensional Planck cell
of volume $L_{\rm planck}^d$, the number of web cells is
\begin{equation}
\frac{\k_d V_0}{L_{\rm planck}^d}
\;\sim\;
\frac{\k_D V_D}{L_{\rm planck}^D}.
\end{equation}
Thus the microscopic counting of wormhole webs is performed in $d$ 
dimensions, whereas the resulting cosmological constant 
\versionII{governs the effective}
$D$--dimensional macroscopic spacetime.

Using \rf{KnittingCC} we obtain
\begin{equation}\label{WormholeWebEnergy}
\frac{E_{\rm wh}}{E_{\rm planck}}
\,=\,
\frac{\Lambda L_{\rm wh}^D / \k_D}{E_{\rm planck}}
\,\sim\,
\frac{\Lambda L_{\rm planck}^D / L_{\rm planck}^{D-1}}{L_{\rm planck}^{-1}}
\,\sim\,
\mu_{\rm wh} L_{\rm planck}^2
A_\ast
\,\sim\,
A_\ast ,
\end{equation}
where $L_{\rm wh}^D/\k_D$ is the $D$--dimensional Planck cell volume,
and $E_{\rm wh}$ denotes the vacuum energy stored in a single
Planck--scale wormhole web, i.e.\ the energy carried by one such
$d$--dimensional cell in the knitted spacetime.
In the last step of \rf{WormholeWebEnergy} we used \rf{WormholeSize}.

Because a cosmological term does not obey an energy conservation law, 
the vacuum energy of the Universe is not conserved. 
In the present framework 
the vacuum energy generated by the wormhole web 
produces a cosmological term whose total contribution 
to the Universe at a given time $t$ is
\begin{equation}\label{VacuumEnergy}
E_\Lambda(t)=\frac{\k_D V_D(t)}{L_{\rm planck}^D}\,E_{\rm wh},
\end{equation}
which grows in proportion to the macroscopic spatial volume $V_D(t)$.

Within the Coleman mechanism,
$\Lambda$ is a dynamical variable whose probability distribution 
is sharply peaked at $\Lambda=0$.
Physically, this implies that at some moment in the early Universe 
the vacuum energy $E_\Lambda$ is removed from the gravitational sector 
and converted into ordinary matter and radiation. 

We assume that this Coleman selection occurs 
immediately after the knitted spacetime has formed, 
when the emergent $D$--dimensional Universe has Planckian volume. 
At the moment when the vacuum selection takes place, 
the spatial volume therefore satisfies 
\begin{equation}\label{PlanckVolume}
  V_D(t_{\rm CM}) \;\sim\; \frac{1}{\k_D} L_{\rm planck}^D ,
\end{equation}
where $t_{\rm CM}$ is the time when the Coleman mechanism occurred.

The vacuum energy of the Universe at that instant, 
given by \rf{VacuumEnergy},
is then deposited into this Planck--scale region
and becomes the big--bang energy of the Universe,
which we define as
\begin{equation}\label{MatterEnergy}
  E_{\rm univ} \define E_\Lambda(t_{\rm CM})
  \sim E_{\rm wh}
  ,
\end{equation}
where in the last step we used \rf{VacuumEnergy} and \rf{PlanckVolume}.

Since $L_{\rm planck}\define L_{\rm wh}\sim1/\sqrt{\mu_{\rm wh}}$,
so that $\mu_{\rm wh}L_{\rm planck}^2\sim1$ independently of the Coleman
mechanism, the hierarchy
\[
  \frac{E_{\rm univ}}{E_{\rm planck}}
  \sim 
  \frac{E_{\rm wh}}{E_{\rm planck}}
  \sim A_\ast
\]
is generated entirely by the wormhole--web amplitude.
Remarkably, this ratio is independent of the emergent spatial dimension $D$;
it is fixed solely by the underlying two--dimensional dynamics
that knits one--dimensional universes into a wormhole web.

\subsection{The ``discretized'' Structure of Spacetime}
\label{sec:DiscretizedSpaceTime}

In the framework of the knitting mechanism, spacetime as a smooth 
higher-dimensional manifold emerges only as an effective description 
of a more microscopic structure, namely the ``wormhole web'' 
that knits together the underlying one-dimensional universes, 
as depicted in Fig.\ \ref{fig2}.
This continuum description is therefore valid only down to the scale 
at which the individual one-dimensional building blocks, connected by 
wormhole cylinders, can no longer be resolved 
by the effective higher-dimensional description.

The typical spatial and temporal sizes of the one-dimensional universes 
are fixed by the one-dimensional cosmological constants that arise 
when the $W_3$ symmetry is broken.
In particular, the corresponding CDT wormhole dynamics implies 
a characteristic length scale \rf{WormholeSize}, 
which, as argued in the previous subsection, 
is dynamically generated by the wormhole-web sector.
Through the knitting process this scale becomes the fundamental 
resolution scale of the emergent higher-dimensional spacetime, 
and we identify it with the Planck length \rf{WormholeWebSize}. 

Since the class of expanding one-dimensional universes reaches 
infinite size within a finite proper time 
measured in the one-dimensional universes 
and set by their cosmological constants, 
the entire knitting process that produces the higher-dimensional spacetime 
necessarily takes place within spacetime regions of Planckian size. 
Consequently, the wormhole web not only generates 
the large-scale geometry but also enforces a minimal length 
and time scale in the emergent spacetime. 

In this sense, the coupling constant $g$ 
plays the role of an order parameter for the knitting mechanism. 
Before knitting, when the one–dimensional universes are 
essentially uncorrelated, $g$ is of order unity in Planck units.
After the wormhole web condenses and a higher–dimensional spacetime emerges, 
$g$ is dynamically driven to an extremely small value.
The smallness of $g$ therefore does not represent fine tuning, 
but signals that spacetime is in a condensed phase of the wormhole–web system.

In this picture the gravitational constant $G_{\rm N}$ 
is not a fundamental input parameter but a derived quantity, 
determined by the wormhole-web dynamics and the associated Planck scale.
It enters the effective minisuperspace Hamiltonian, 
such as Eq.\ \rf{g2}, only after the knitting has taken place, 
and its magnitude is fixed by the same mechanism 
that produces the discretized structure of spacetime 
at the scale $L_{\rm planck}$.


\subsection{On the Low-Entropy State of the Early Universe}
\label{sec:LowEntropy}

In the scenario proposed by our theory, 
the universe did not originate from a thermal fluctuation 
but emerged from ``nothing'' as multiple flavored point-like configurations, 
which subsequently expanded into one-dimensional loops.
It then immediately entered a phase of inflation, 
during which the universe expanded as a one-dimensional spatial entity. 
When the cosmic scale is  getting close to the Planck length, 
the knitting mechanism will operate and  trigger a transition 
to a higher-dimensional universe. 
In this process, 
the vacuum energy is converted into the energy of matter fields 
through the Coleman mechanism, marking the end of inflation.

During the initial inflationary phase the universe is 
effectively one-dimensional, 
and transverse gravitational degrees of freedom do not exist.
In this reduced system the homogeneous expanding state 
is a typical high-entropy configuration.

When the knitting mechanism triggers the transition 
to a higher-dimensional spacetime, new gravitational degrees of freedom 
suddenly become available. 
This does not reduce the entropy of the Universe. 
Instead, it dramatically enlarges the space of accessible 
gravitational microstates, and thereby increases the \textit{maximal} 
possible entropy. 
As a result, the actual state of the Universe—although 
continuously evolving with increasing entropy—now occupies 
an extremely small fraction of the newly opened phase space.
From the viewpoint of the higher-dimensional gravitational dynamics, 
the Universe therefore appears to be in a low-entropy state. 

In this way the Penrose low-entropy condition is not imposed by hand, 
but arises dynamically from the sudden increase of dimensionality 
induced by the knitting mechanism.

\subsection{Scale-Invariant Spectrum}
\label{sec:ScaleInvariance}

In our model the very early Universe is described by 
the expansion of one--dimensional spaces governed by 
the $W_3$--based dynamics and the tangent--type solution \rf{e15}.
These one--dimensional universes later become knitted together, 
forming the higher--dimensional spacetime in which we live.
The primordial fluctuations therefore originate from 
quantum fluctuations of the flavored one--dimensional universes 
before knitting, rather than from an ordinary scalar field 
in a two--dimensional Robertson--Walker spacetime.

Let $f_k$ denote a generic scalar perturbation mode 
associated with a given flavor 
(one of the individual one--dimensional universes) 
before knitting. 
Because the dynamics is effectively one--dimensional, 
its quantum fluctuations obey
\begin{equation}
\langle |\delta f_k|^2 \rangle \;\propto\; \frac{1}{k},
\label{1Dfluct}
\end{equation}
which follows from mode normalization and canonical commutation relations,
and is independent of the detailed form of the background scale factor 
$\ScF(t)$.

The knitting mechanism transforms each one--dimensional mode $f_k$ 
into a perturbation of the higher--dimensional geometry, 
effectively distributing the original $1/k$ scaling 
across the emergent spacetime. 
Consequently, the power spectrum in the emergent cosmology is
\begin{equation}
P(k) \;\define\; k \, \langle |\delta f_k|^2 \rangle \;=\; \text{const},
\end{equation}
corresponding to a scale--invariant spectrum with spectral index
\begin{equation}
n_s \simeq 1 .
\end{equation}

This scale invariance arises fundamentally differently 
from standard three--dimensional inflationary cosmology, 
where it originates from horizon crossing in a near--de~Sitter expansion.
Here, it stems directly from the one--dimensional nature 
of the microscopic degrees of freedom before knitting, 
with the tangent--type expansion \rf{e15} controlling the background evolution.
\versionII{
A key feature of the knitting mechanism is that 
the formation of wormholes occurs at the Planck scale 
and takes place globally across the universe. 
Because the transition is global, 
it does not introduce relative distortions between different regions. 
In this sense, it is natural to expect that 
the scaling structure present before the transition is preserved 
at the macroscopic level, at least to a good approximation. 
This is fundamentally different from scenarios based on 
inflaton potentials, 
where the dynamics is typically local and field-dependent. 
A more rigorous derivation of this property from the microscopic dynamics 
is left for future work, 
but if it works out we can say that the knitting of 
one–dimensional universes into a higher–dimensional spacetime 
not only generates the geometry of our Universe, 
but also naturally explains the observed near scale–invariance 
of primordial fluctuations. 
However, since the deviation from scale-invariance is measured 
with great precision, it is clear that to have any value, 
such an explanation has to be able to reproduce 
this deviation with great precision.
}

\section{Summary}\label{summary}

We have developed a multiverse model based on 
a multicomponent $W_3$ algebra.
In this model, symmetry breaking first leads to 
the appearance of time and subsequently 
to one-dimensional universes carrying different ``flavors''.
These one-dimensional universes are created as point-like objects, 
without spatial extension, 
thereby avoiding any conventional ``initial'' singularity.
They subsequently expand according to Eq.\ \rf{e15}, 
but also interact and become knitted together by wormholes, 
forming a wormhole web,
where the wormholes themselves are one-dimensional universes
carrying their own flavors.
Through this knitting process, 
a higher-dimensional spacetime emerges 
as an effective description at a characteristic scale 
set by the wormhole sector, 
a scale that we identify with the Planck length.

It is important to emphasize that in this framework
the cosmological constant does not primarily represent
a classical energy density of spacetime,
but rather the quantum weight of microscopic spacetime configurations.
In a quantum-gravitational setting,
a large cosmological constant corresponds to a large statistical weight
assigned to certain microscopic geometries,
rather than to a large macroscopic curvature.
This shift in interpretation is crucial for understanding
how Planckian vacuum energy generated by the wormhole web
can coexist with an effectively vanishing cosmological constant
in the emergent classical universe.
More precisely, the relation $\Lambda \sim \mu_{\rm wh} A_\ast$ 
shows that the macroscopic cosmological constant 
is the statistical imprint of Planck--scale wormhole webs 
in the knitted spacetime.

Above this dynamically generated Planck scale, 
we assume that the Coleman mechanism drives 
the ordinary cosmological constant to zero, 
converting the corresponding vacuum energy 
into matter degrees of freedom. 
\versionII{%
Since the Coleman mechanism is expected to become effective
at roughly the same scale at which the wormhole knitting occurs,
the large cosmological constant generated by the knitting process
may be rapidly relaxed.
This raises the possibility that the associated inflationary expansion 
remains compatible with cosmological observations, 
although at present we do not know how to determine 
the corresponding e-folding number within the present framework.}
The resulting effective higher-dimensional minisuperspace Hamiltonian,
Eq.\ \rf{g2}, leads to a modified Friedmann equation. 
Compared to the standard Friedmann equation, 
it contains an additional term with coupling constant $g$, 
which admits an interpretation in terms of baby universes 
being absorbed into our Universe. 
This process induces a late-time accelerated expansion 
without the need for an explicit cosmological constant, 
which is assumed to vanish by the Coleman mechanism.
We have argued that this late-time dynamics allows for 
a resolution of the observed $H_0$ tension.

The resulting effective coupling is of order
$g \approx \versionII{10^{-183}} g_0$, 
where $g_0$ is the ``bare'' coupling constant before knitting.
This value is consistent with that required to fit 
cosmological observations, supporting the idea that 
the knitting mechanism can generate the observed smallness of $g$. 

The emergence of a macroscopic and effectively classical spacetime 
is governed by a phase transition controlled by the coupling $g$, 
which plays the role of an order parameter for the knitting mechanism. 
In the small--$g$ phase, the interaction between different universes 
is strongly suppressed, and the Universe evolves according to 
classical gravitational dynamics.

The modified Friedmann equation defines a special ``singular'' point 
when its evolution curve is viewed as an algebraic curve.
This point introduces a natural scale that can account for 
several large dimensionless ratios observed in cosmology.
Furthermore, the model offers the possibility of explaining 
both the low-entropy initial state and the nearly scale-invariant 
spectrum of primordial fluctuations.

Central to this framework are the knitting and exchange mechanisms 
mediated by wormholes, which generate a higher-dimensional condensate 
from the one-dimensional universes of different flavors.
We have argued that this condensate may be described 
as a $\tau$-function subject to generalized Virasoro constraints.
A major open goal is to construct this $\tau$-function explicitly, 
which would allow many of the conjectural aspects of the model 
to be placed on a more rigorous footing.

\vspace{1cm}
\noindent{\textbf{\Large Acknowledgements}}\\

This work was supported by JSPS KAKENHI Grant Numbers JP25K07278.

\end{document}